\documentclass[twocolumn]{aastex62}
\usepackage{xfrac}

\received{\today}
\revised{date}
\accepted{date}

\submitjournal{ApJ}

\shorttitle{Post-outburst life of V346 Nor}
\shortauthors{K\'osp\'al et al.}

\begin{document}

\title{V346 Nor: the post-outburst life of a peculiar young eruptive star}


\author[0000-0001-7157-6275]{\'A. K\'osp\'al}
\affiliation{Konkoly Observatory, Research Centre for Astronomy and Earth Sciences, Konkoly-Thege M. \'ut 15-17, 1121 Budapest, Hungary}
\affiliation{Max Planck Institute for Astronomy, K\"onigstuhl 17, D-69117 Heidelberg, Germany}

\author[0000-0001-9830-3509]{Zs. M. Szab\'o}
\affiliation{Konkoly Observatory, Research Centre for Astronomy and Earth Sciences, Konkoly-Thege M. \'ut 15-17, 1121 Budapest, Hungary}
\affiliation{E\"otv\"os Lor\'and University, Department of Astronomy, P\'azm\'any P\'eter s\'et\'any 1/A, 1117 Budapest, Hungary}

\author[0000-0001-6015-646X]{P. \'Abrah\'am}
\affil{Konkoly Observatory, Research Centre for Astronomy and Earth Sciences, Konkoly-Thege M. \'ut 15-17, 1121 Budapest, Hungary}

\author[0000-0001-6017-8773]{S. Kraus}
\affiliation{School of Physics, University of Exeter, Stocker Road, Exeter EX4 4QL, UK}

\author[0000-0001-9248-7546]{M. Takami}
\affiliation{Institute of Astronomy and Astrophysics, Academia Sinica, 11F of Astronomy-Mathematics Building, AS/NTU No.1, Sec. 4, Roosevelt
Rd, Taipei 10617, Taiwan, R.O.C.}

\author{P. W. Lucas}
\affiliation{Centre for Astrophysics, University of Hertfordshire, College Lane, Hatfield AL10 9AB, UK}

\author{C. Contreras Pe\~na}
\affiliation{School of Physics, University of Exeter, Stocker Road, Exeter EX4 4QL, UK}

\author[0000-0001-5207-5619]{A. Udalski}
\affiliation{Astronomical Observatory, University
of Warsaw, Al. Ujazdowskie 4, 00-478 Warszawa, Poland}


\begin{abstract}
FU Orionis-type objects (FUors) are young low-mass stars undergoing powerful accretion outbursts. The increased accretion is often accompanied by  collimated jets and energetic, large-scale molecular outflows. The extra heating during the outburst may also induce detectable geometrical, chemical, and mineralogical changes in the circumstellar material, affecting possible planet formation around these objects. V346 Nor is a southern FUor with peculiar spectral characteristics. Decades after the beginning of its outburst, it unexpectedly underwent a fading event around 2010 due to a decrease in the mass accretion rate onto the star by at least two orders of magnitude. Here we present optical and near-infrared photometry and spectroscopy obtained after the minimum. Our light curves show a gradual re-brightening of V346~Nor, with its $K_{\rm s}$-band brightness only 1.5\,mag below the outburst brightness level. Our VLT/XSHOOTER spectroscopic observations display several strong forbidden emission lines towards the source from various metals and molecular hydrogen, suggesting the launch of a new jet. Our $N$-band spectrum obtained with VLT/VISIR outlines a deeper silicate absorption feature than before, indicating that the geometry of the circumstellar medium has changed in the post-outburst period compared to peak brightness.
\end{abstract}



\section{Introduction} \label{sec:intro}

Mass accretion from the circumstellar disk onto the protostar is a fundamental process during star formation. Young stars typically reach their final mass within a few million years of the start of the gravitational collapse \citep[e.g.,][]{dunham2014ppvi}. During this time, the accretion rate gradually decreases. However, there are observational and theoretical \citep{vorobyov2006,vorobyov2010,vorobyov2013} evidence showing that accretion is not a smooth process but highly variable. The largest accretion rate changes are observed in the FU~Orionis-type objects (FUors), which are identified based on their unpredictable brightenings of several magnitudes observed at optical or near-infrared wavelengths \citep{hk96, audard2014}. Such FUor-type outbursts are attributed to a highly enhanced accretion rate. Assuming that all young stars undergo eruptive phases, the FUor-phenomenon is suggested to solve the protostellar luminosity problem, i.e.~that young stars are observed to be about an order of magnitude fainter (i.e., accrete less) than theoretically predicted \citep{dunham2013}.

\begin{figure}[ht!]
\includegraphics[width=\columnwidth]{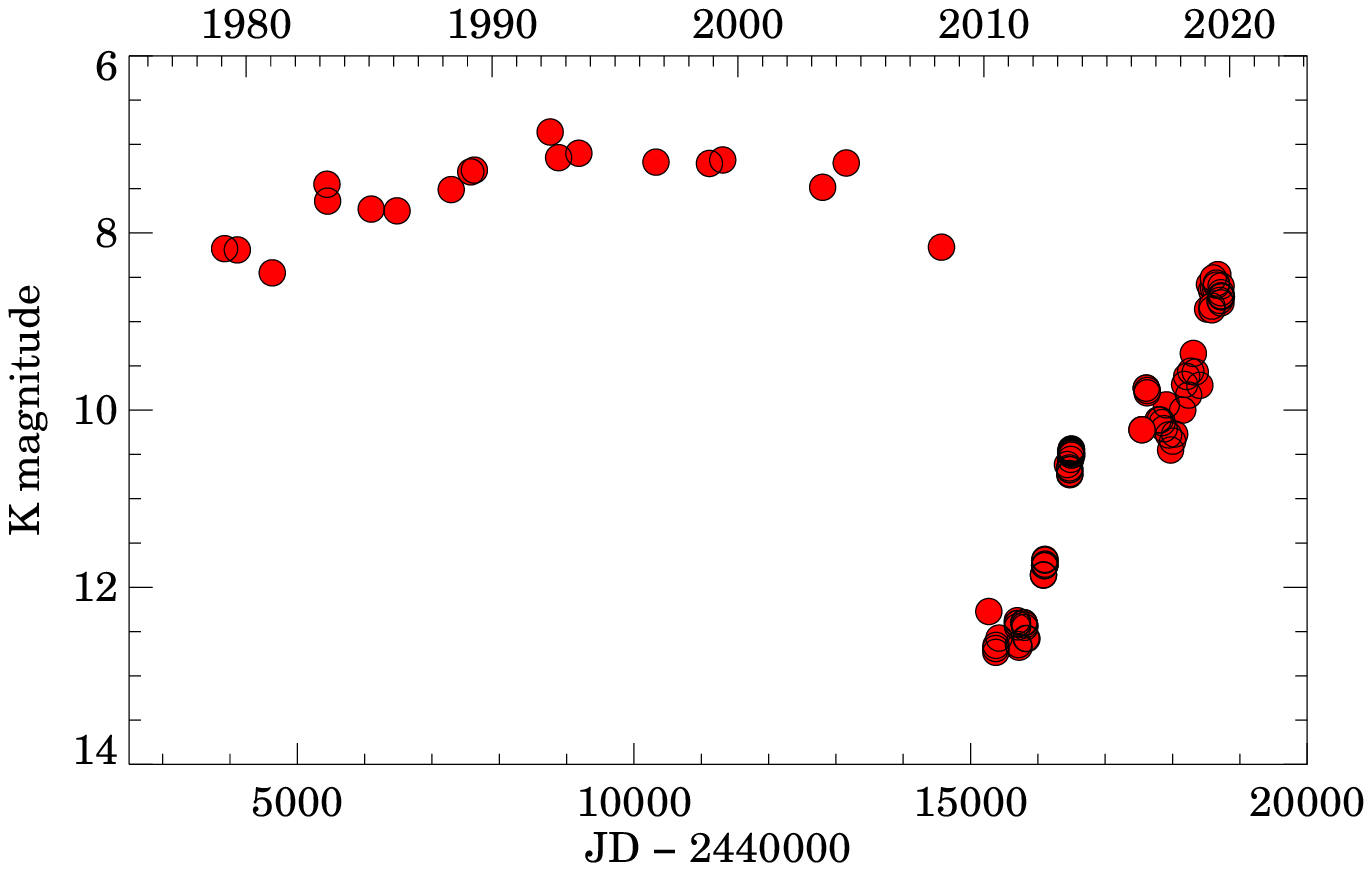}
\caption{$K$-band light curve of V346~Nor. Data are from \citet{elias1980,reipurth1983b,reipurth1985,graham1985,reipurth1991,molinari1993,prusti1993,reipurth1997,kraus2016,kospal2017a,connelley2018}.\label{fig:kmag}}
\end{figure}

V346~Nor is a FUor in the southern hemisphere, which went into outburst around 1980 \citep[][and references therein]{kospal2017a}. It is located in the Sa~187 molecular cloud at a distance of 700\,pc. It is the driving source of the Herbig--Haro object HH~57, at a distance of about 8$''$ \citep{graham1985}. Although it clearly showed an accretion outburst, V346~Nor is a peculiar FUor in several aspects that are traditionally considered as classification criteria. It shows no broad water absorption bands in the near-infrared and no CO overtone absorption in the $K$ band. On the other hand it shows emission lines that are not present in bona fide FUors \citep{reipurth1985, graham1985, connelley2018}. The $K$-band photometry of V346~Nor in Fig.~\ref{fig:kmag} shows that its light curve is also special: after almost 30 years in the bright outburst state, it displayed a sudden fading with a deep minimum around 2010-11, which we attributed to a significant drop in the accretion rate \citep{kraus2016,kospal2017a}.

Although V346~Nor started brightening again almost immediately after its minimum, this curious temporary stop in the accretion rate motivated us to study the system further and to find out what physical changes happened since 2010-11 (hereafter referred to as the post-outburst state). Here, we report on new near-infrared and mid-infrared spectroscopy from 2015-16 and present our optical and near-infrared photometric monitoring for V346~Nor. In Sec.~\ref{sec:obs} we describe the observations, in Sec.~\ref{sec:res} we show our photometric and spectroscopic results, in Sec.~\ref{sec:spectral} we analyze the spectra, in Sec.~\ref{sec:disc} we discuss the implications of our observations, and in Sec.~\ref{sec:conclusions} we summarize our conclusions.

\section{Observations and Data Reduction}
\label{sec:obs}

\subsection{Optical and Infrared Photometry}

We monitored V346 Nor with the SMARTS 1.3\,m telescope at Cerro Tololo (Chile) between 2016 June 6 and 2019 July 9 with an approximately monthly cadence. We used the ANDICAM instrument, that makes simultaneous optical (with a 2k$\times$2k Fairchild 447 CCD) and infrared (with a Rockwell 1024$\times$1024 Hawaii detector) exposures. On each observing night we typically obtained 5 images in the $I$ band, and 5--10 images in the $J$, $H$, and $K_{\rm s}$ bands. More information on the instrument, data reduction, and photometry can be found in \citet{kospal2017a}, where the results for the first two epochs are also presented. These numbers, together with our subsequent monitoring are  given in Tab.~\ref{tab:phot}. In the $I$ band, V346 Nor was undetected in 2016--2017. For these observations, we determined upper limits for its magnitude by computing the average 3$\sigma$ noise in the background, and converting it to standard magnitude using the relationship between the instrumental and standard magnitude derived from the comparison stars. These upper limits are also listed in Tab.~\ref{tab:phot}.

We could identify V346~Nor in the archive of the Optical Gravitational Lensing Experiment \citep[OGLE,][]{udalski2015}  under the name gd1112.05.9732. We collected all photometric observations obtained since 2010 in the $I_{\rm C}$ band with the 1.3\,m Warsaw Telescope at Las Campanas Observatory in Chile. The telescope is equipped with a 256 Megapixel mosaic CCD camera with a field of view of 1.4 square degrees at a scale of 0$\farcs$26/pixel. Details of the OGLE instrumental setup, data reduction and calibration are in \citet{udalski2015}. We found that V346~Nor was not detected in 2010--2012, indicating that it was fainter than $I_{\rm C}\sim22$\,mag. The source became visible after 2013 in frames taken with 25--30\,s exposure time. The photometric results are listed in Tab.~\ref{tab:phot}.

V346~Nor was covered as part of two ESO public near-infrared surveys: the VISTA Variables in the Via Lactea survey (VVV) and its extension VVVx. Both surveys used the VISTA 4.1\,m telescope at Paranal Observatory in Chile and the VIRCAM near-IR camera \citep{minniti2010}. We downloaded all VIRCAM images that include V346~Nor. To obtain photometry, we performed our own flux extraction as described in \citet{kospal2017a}. As an important step, in the $K_{\rm s}$ frames we performed correction for the known non-linearity of the VIRCAM detectors for bright sources \citep[e.g.,][]{saito2012}. V346~Nor fell into the nonlinear regime. The correction method is described in detail in Appendix A of \citet{kospal2017a}. The $J$ and $H$ photometry, as well as the $K_{\rm s}$ band results after the non-linearity correction are given in Tab.~\ref{tab:phot}.

At mid-infrared wavelengths we downloaded all time-resolved observations from the ALLWISE and NEOWISE-R Single Exposure Source Table in the W1 (3.4$\,\mu$m) and W2 (4.6$\,\mu$m) photometric bands, and computed their seasonal averages after removing the outlier data points \citep{wright2010,mainzer2014}. In the error budget, we added to the photometric uncertainties in quadrature 2.4\% and 2.8\% as the uncertainty of the absolute calibration in the W1 and W2 bands, respectively \citep[Sect.~4.4 of the WISE Explanatory Supplement,][]{cutri2015}.

\begin{figure*}
\includegraphics[width=\textwidth]{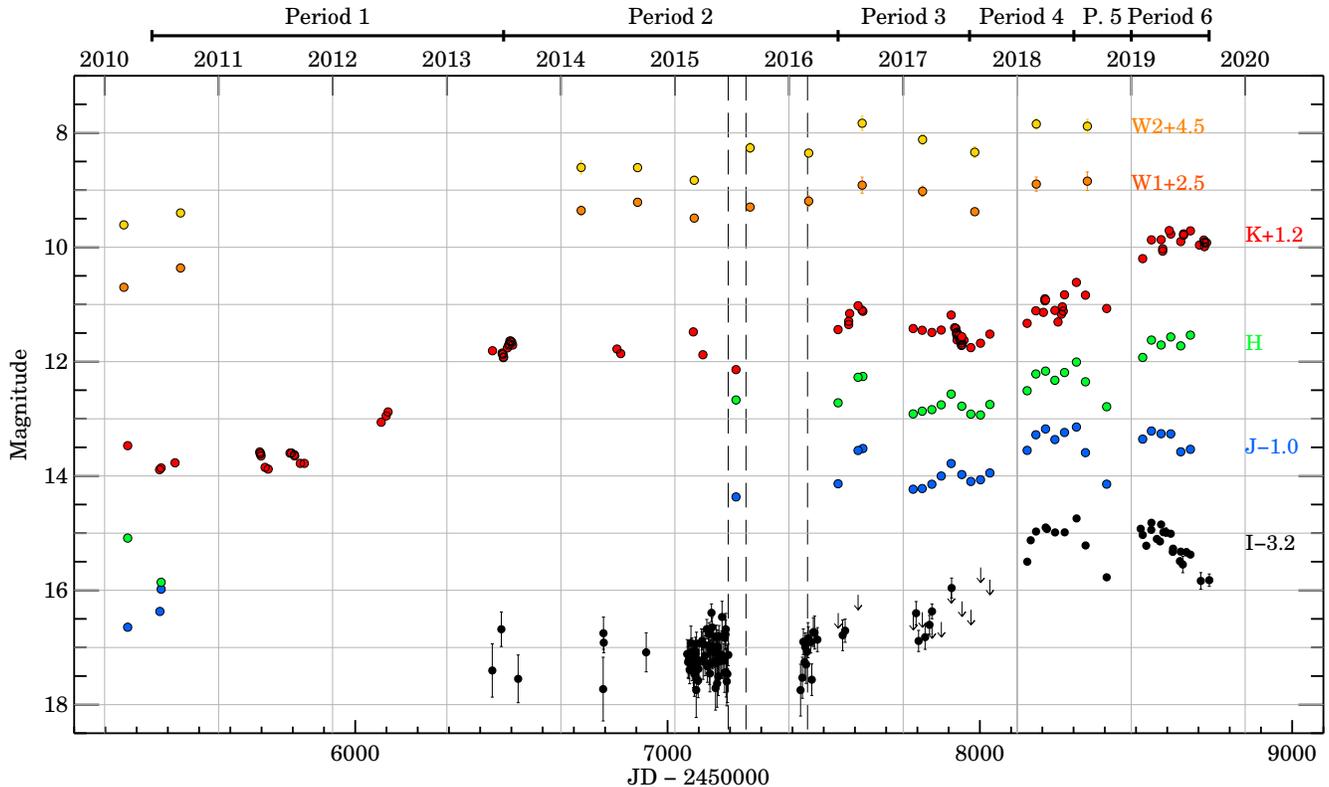}
\caption{Optical and infrared light curves of V346~Nor. Data are from \citet{kraus2016}, \citet{kospal2017a}, and this paper (Tab.~\ref{tab:phot}). For clarity, the light curves were shifted along the y axis by the indicated magnitudes. The vertical lines in 2015 mark when our XSHOOTER spectra were taken, the vertical line in 2016 indicates the date of our VISIR spectrum. Error bars smaller than the symbol size are not plotted. The periods above the graph refer to those discussed in Sec.~\ref{sec:lightcurveevolution}.\label{fig:lightcurve}}
\end{figure*}

\begin{figure*}
\begin{center}
\includegraphics[width=0.75\textwidth]{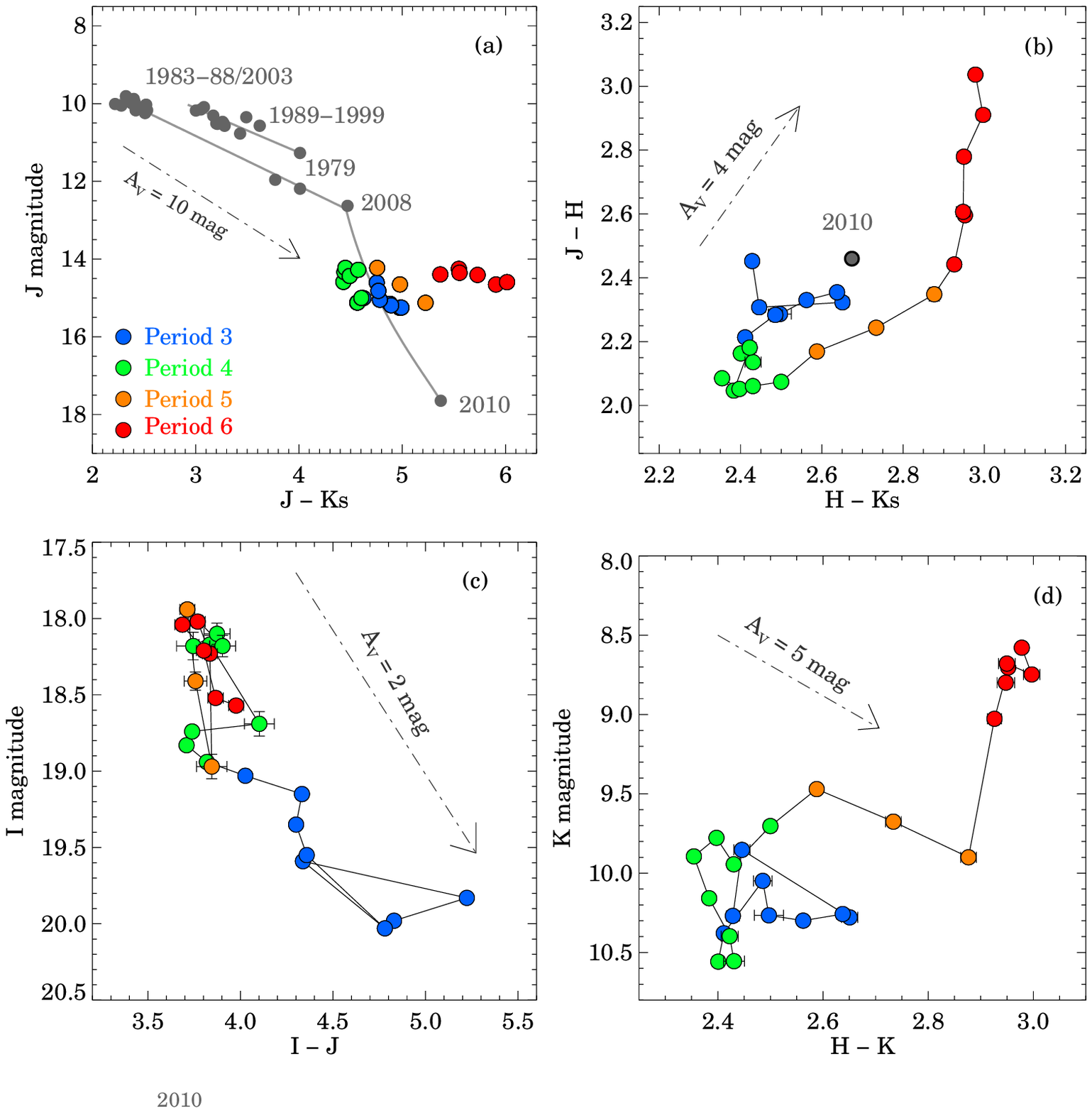}
\caption{Optical and near-infrared color--color and color--magnitude diagrams of V346~Nor. Error bars smaller than the symbol size are not plotted.\label{fig:tcd}}
\end{center}
\end{figure*}

We used the photometric results summarized above to construct the optical and infrared light curves, as well as color-color and color-magnitude diagrams for V346~Nor, which can be seen in Figs.~\ref{fig:kmag}, \ref{fig:lightcurve}, and \ref{fig:tcd}.

\subsection{Near-Infrared Spectroscopy}

We took spectra of V346~Nor on 2015 June 20, July 12, and July 16 using the XSHOOTER \'echelle spectrograph \citep{vernet2011} on ESO's Very Large Telescope (VLT) (095.C-0765, PI: S.~Kraus).  We used the 1$\farcs$0, 0$\farcs$9, and 0$\farcs$4 wide slits, providing 5400, 8900, and 11600 spectral resolution in the UVB (0.30--0.56$\,\mu$m), VIS (0.53--1.02$\,\mu$m), and NIR (0.99--2.48$\,\mu$m) arms, respectively. We downloaded the raw data and the necessary calibration files from the ESO Science Archive  Facility. On the first night, V346 Nor was positioned in the center of the slit, while on the other two nights, an ABBA nodding pattern was applied. In addition, on the second night, one observation was executed with the 5$''$ wide slit to enable correction for slit losses. We reduced the raw data using the X-SHOOTER pipeline v.3.3.5 within the EsoReflex v.2.9.1 environment. We performed standard data reduction, but set the sky regions manually, in order to avoid emission from HH~57 and a nearby red point source in the slit. To extract the 1D spectrum of V346~Nor, we used an aperture with a width between 1$\farcs$4 and 2$\farcs$2, depending on the measured diameter of the object in the spatial direction.

\begin{figure*}
\includegraphics[width=\textwidth]{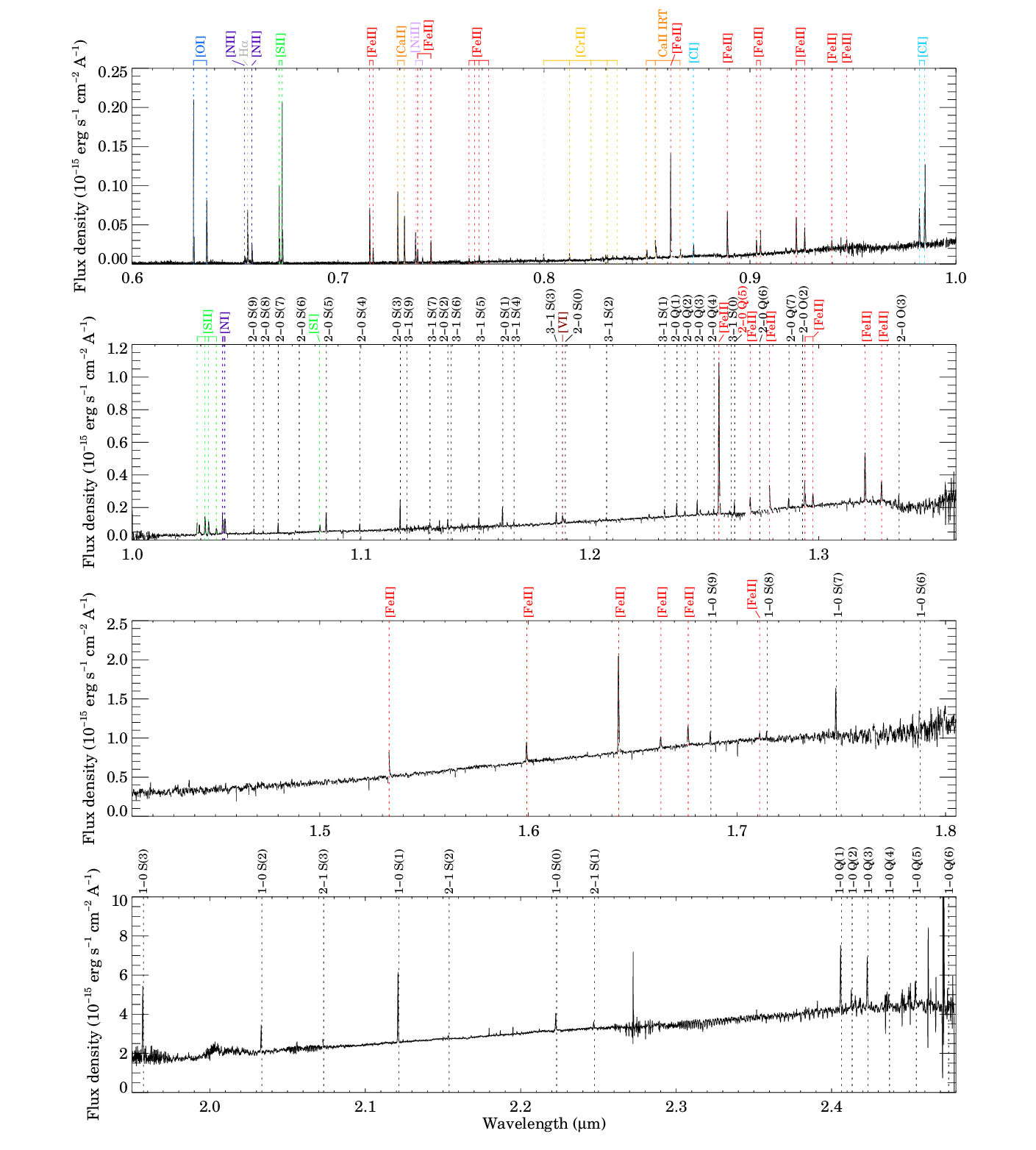}
\caption{Average VLT/XSHOOTER spectrum of V346~Nor observed in 2015 June-July. Note the different y axis scales. The top panel shows the visual spectrum, while the bottom three panels show those parts of the near-infrared spectrum  that are not affected by severe telluric absorption. The identified lines are marked by red vertical dotted lines and the corresponding species (or transition for H$_2$) are marked above each panel. \label{fig:xshooter}}
\end{figure*}

We corrected for telluric absorption using ESO's molecfit tool \citep{smette2015, kausch2015}. Then we scaled each spectrum so that their continuum level matched those observed with the 5$''$ broad slit. This could only be done for the VIS and NIR arm, because V346~Nor was undetected in the UVB arm. We noticed and corrected for a slight systematic shift in the wavelength of emission lines in V346 Nor between the three different observing nights. Finally, because we did not see significant variability in any of the emission lines, we averaged all spectra to increase the signal-to-noise ratio. The resulting spectrum can be seen in Fig.~\ref{fig:xshooter}.

\subsection{Mid-Infrared Spectroscopy}
\label{sec:visirobs}

\begin{figure}
\includegraphics[width=\columnwidth]{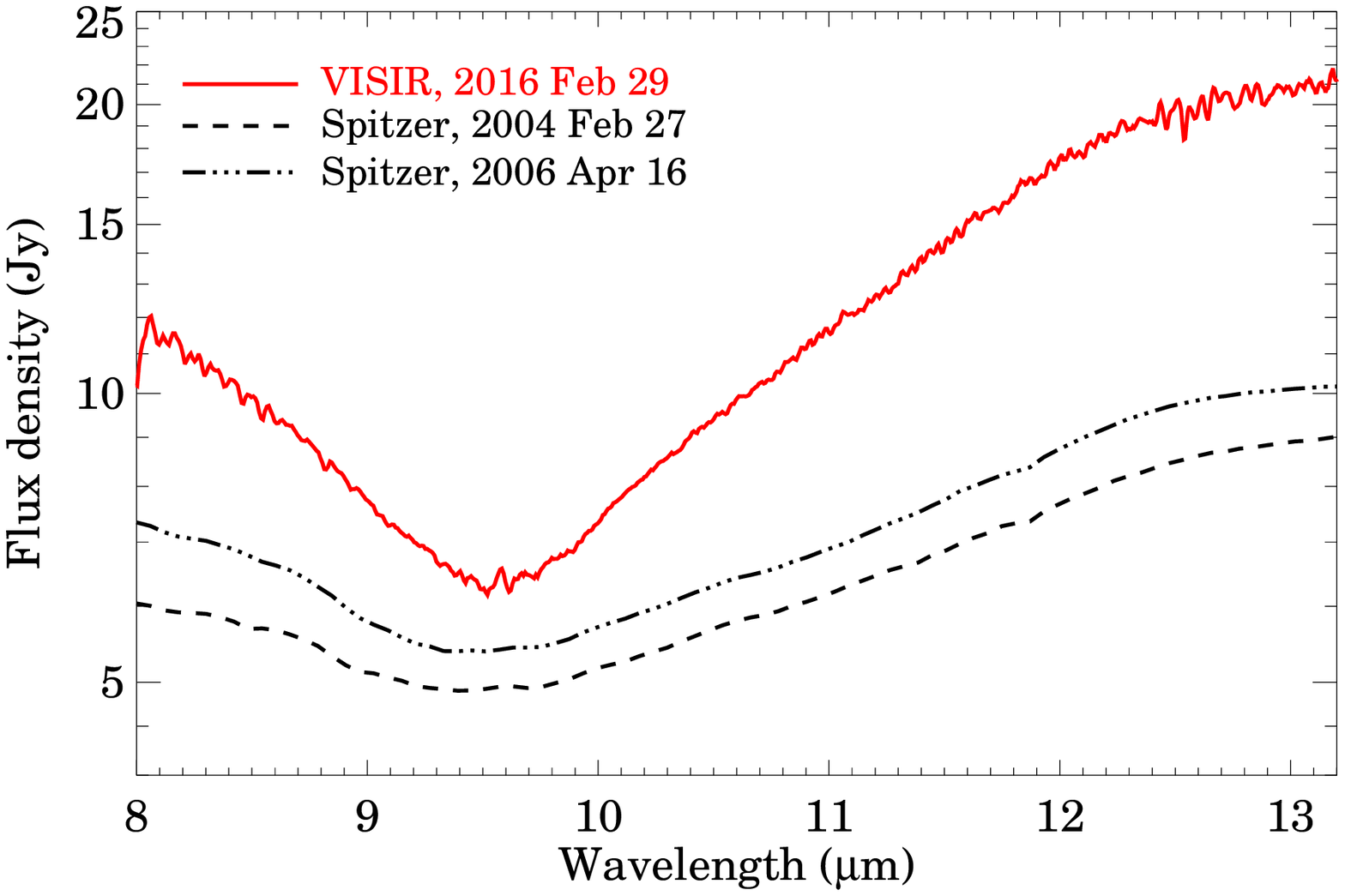}
\caption{VLT/VISIR $N$-band spectrum of V346~Nor, obtained in the post-outburst phase (red). For comparison, we  plotted two earlier Spitzer/IRS measurements from the outburst state (black). \label{fig:visir}}
\end{figure}

We obtained an $N$-band spectrum of V346~Nor using VISIR \citep{lagage2004} on the VLT (095.C-0765, PI: S.~Kraus). The observations were performed on 2016 February 29 with an exposure time of 2300\,s, and a slit size of $0{\farcs}4$, which provided a spectral resolution of $R\sim 350$. We observed HD~149447 and HD~151680 as telluric and flux standard stars, whose different zenith distances allowed us to correct for airmass differences. For basic data reduction and extraction of the spectra we run the ESO VISIR spectroscopic pipeline in the EsoRex environment version 3.12.3. In order to correct for telluric features, we divided the target spectrum by a spectrum derived via interpolation between the two standard star measurements, one obtained at higher and one at lower airmass than V346 Nor. Flux calibration was carried out by multiplying by the model spectra of the standard stars. The resulting spectrum is presented in Fig. \ref{fig:visir}.

\section{Results}
\label{sec:res}

\subsection{Light curves and color evolution}
\label{sec:colorevolution}

The brightness and color evolution of V346~Nor for the period 2010--2019 are shown in Figs.~\ref{fig:lightcurve} and \ref{fig:tcd}, respectively. Following the 2010 minimum, a gradual brightening happened until 2013, with periodic modulations superimposed \citep{kraus2016,kospal2017a}. Then the source was constant in 2013--2015 within about $\pm$0.3\,mag. This state can be represented by the following average magnitudes: $I$ = 20.32,\,mag, $J$ = 15.36\,mag, $H$ = 12.67\,mag, $K_{\rm s}$ = 10.84\,mag, $W1$ = 6.83\,mag, $W2$ = 4.07\,mag. Our XSHOOTER spectra were taken during this period. 

\begin{figure}
\includegraphics[width=\columnwidth]{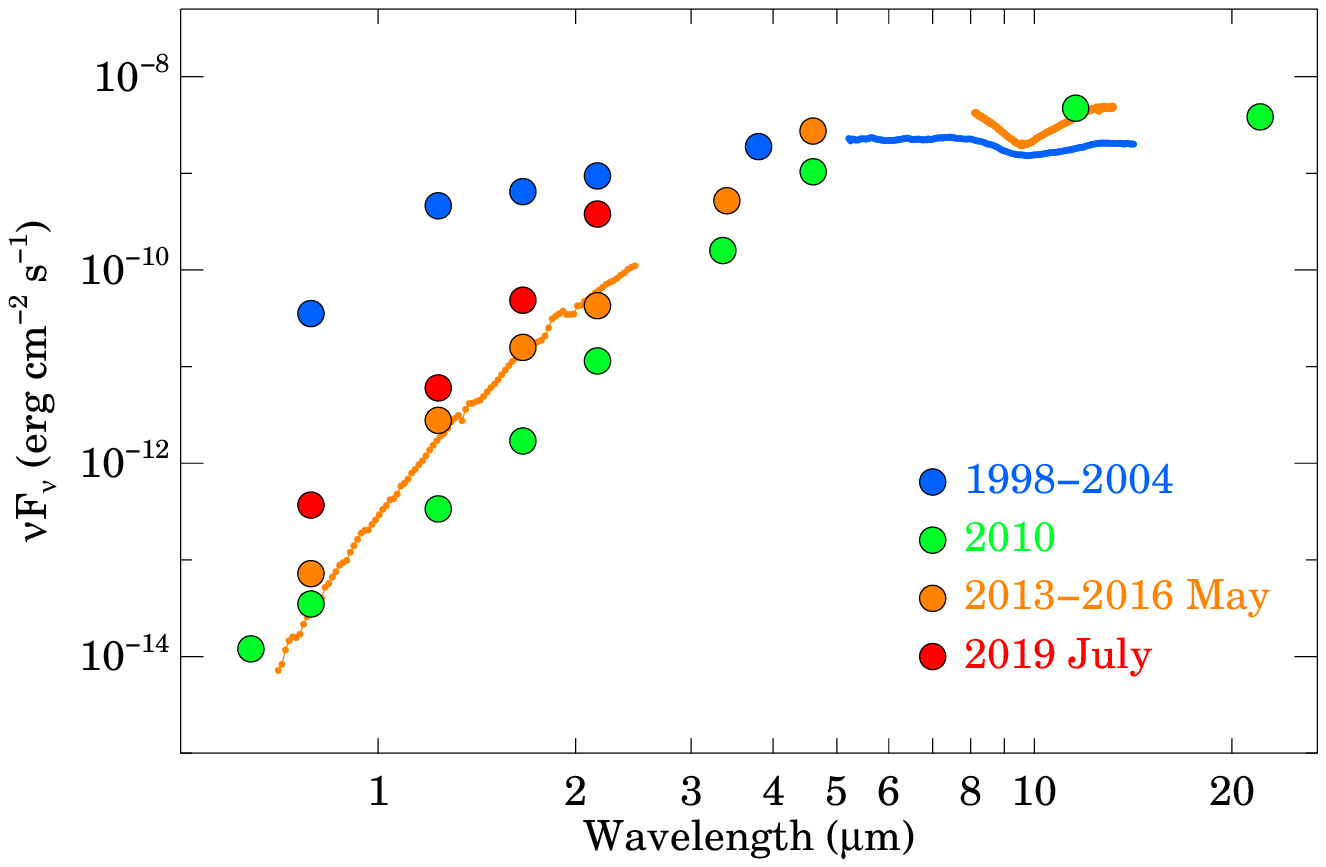}
\caption{SED of V346~Nor at different epochs. Error bars smaller than the symbol size are not plotted.\label{fig:sed}}
\end{figure}

In order to compare quantitatively the 2013--2015 flux levels with the outburst phase, we computed the average magnitudes for the period 1998--2004, when the source was still at maximal brightness, using the photometric data collected by \citet{kospal2017a}. The results are $I$ = 13.59\,mag, $J$ = 9.80\,mag, $H$ = 8.64\,mag, $K$ = 7.34\,mag, and $L$ = 4.98\,mag. Comparing them with the corresponding values from 2013--15 shows a significant brightness drop at all wavelengths: ${\Delta}I$ = 6.7\,mag, ${\Delta}J$ = 5.6\,mag, ${\Delta}H$ = 4.0\,mag, ${\Delta}K_{\rm s}$ = 3.5\,mag, and ${\Delta}L$ = 1.9\,mag. The color changes indicate that the source was much redder in 2013--2015, in the post-outburst phase, than in 1998--2004 and was still significantly fainter than in the outburst state (see also Figs.~\ref{fig:tcd}, \ref{fig:sed}).

From 2016 until 2017 August V346~Nor remained at a similar level as before, apart from two local maxima in 2016 August and 2017 June. The first peak can also be seen in the WISE bands. The peaks' amplitudes do not depend strongly on the wavelength. By 2017 August V346~Nor practically returned to (or slightly exceeded) the average flux level of the 2013--2015 period in all photometric bands. Our VISIR spectrum was obtained at the beginning of this period.

Between 2017 September and 2018 July the main trend was a general flux rise, which led to the brightening of the source by 1--1.5\,mag at all wavelengths. The brightening rate was similar to that in 2011--13, just after the 2010 minimum. Fig.~\ref{fig:tcd} suggests that the brightening was accompanied by a weak blueing. In 2018 August an unexpected rapid fading of the source occurred, with ${\Delta}I=1.03$\,mag, ${\Delta}J=0.85$\,mag, ${\Delta}H=0.7$\,mag, ${\Delta}K=0.36$\,mag over a period of only three months.


By the time we measured V346~Nor again in 2019 February, it became brighter again, and in the $H$ and $K$ bands even exceeded the 2018 level. Interestingly, this time the brightening was accompanied by an unusual color evolution, as the object became redder when brighter. Following a short stable period, since 2019 May V346~Nor has been changing again, in a strongly wavelength dependent way. At shorter wavelengths ($I$ and $J$ bands) it is quickly fading, while at longer wavelengths ($H$ and $K_{\rm s}$ bands) it continues brightening. This introduces a further reddening in the color of the source, thus currently it is even redder than during the 2010 minimum, in spite of the fact that it is orders of magnitudes brighter. If we compare the latest magnitudes from 2019 with those from 2003--2004, it seems that V346~Nor approaches the outburst state in the $K_{\rm s}$ band (currently about 1.5\,mag below, Fig.~\ref{fig:kmag}), but still well below that in the $I$, $J$, and $H$ bands.

\subsection{XSHOOTER spectroscopy}

We detected neither the continuum nor any lines in the UVB arm of our XSHOOTER spectra. Therefore, in the following we will analyze the visual and NIR observations. The continuum is securely detected longward of about 0.75$\,\mu$m, and is steeply rising toward longer wavelengths (Figs.~\ref{fig:xshooter}, \ref{fig:sed}). We detected no absorption lines in V346~Nor, suggesting that the protostellar photosphere was still highly veiled and dominated by emission from the accreting material in 2015-16.

Several emission lines are present in the spectrum of V346~Nor. Almost all of them are shock-related forbidden lines, such as those of [FeII], [NI], [NII], [OI], [SII], [CaII], [NiII], [CI], [CrII], and [VI] (Fig.~\ref{fig:xshooter}). This indicates the presence of a jet in the V346~Nor system. The only permitted emission lines we detected are the H$\alpha$ line of atomic hydrogen and the CaII infrared triplet (IRT). Concerning molecular lines, we detected many H$_2$ rovibrational lines, like the $J$=2--0 S, $J$=2--0 Q, $J$=2--0 O, $J$=2--1 S, $J$=1--0 S, $J$=1--0 Q, and $J$=3--1 S branches, which are often associated with jets in young stellar objects \citep[e.g.,][]{takami2006}. We detect cool CO overtone absorption in the 2.29--2.34$\,\mu$m region. We used the paper of \citet{black1987} to identify the H$_2$ lines and collected $A$ transition probabilities, statistical $g_J$ factors, and $E_J$ upper level energies for them from \citet{turner1977} and \citet{wolniewicz1998}. For the metallic lines, we used the NIST Atomic Spectra Database\footnote{https://www.nist.gov/pml/atomic-spectra-database} to identify them and collect the $A_{ki}$ transition probabilities and $E_i$ and $E_k$ lower and upper level energies. This information is presented in the Appendix in Tabs.~\ref{tab:h2} and \ref{tab:lines}.

\begin{figure}
\includegraphics[width=\columnwidth]{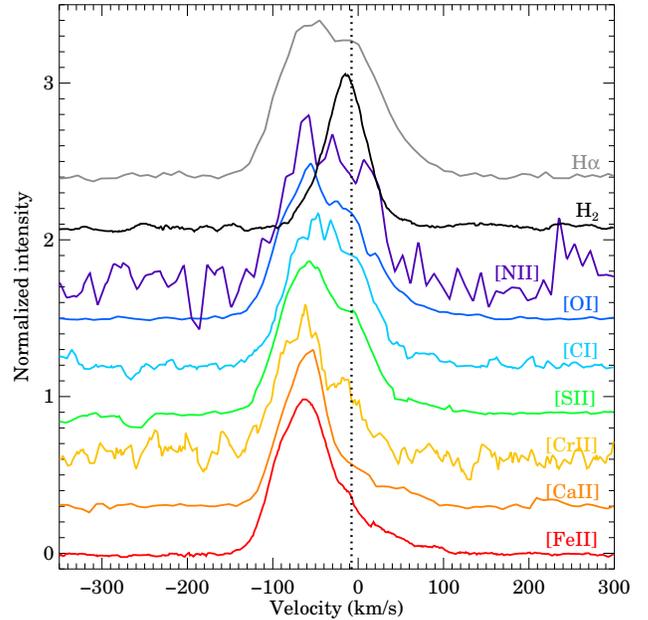}
\caption{Median line profiles for emission lines detected in V346~Nor. [VI] is not plotted because we detected only one line, and [NI] is not plotted because all detected lines are affected by blending. The vertical dotted line marks the heliocentric systemic velocity of $-7.8$\,km\,s$^{-1}$ converted from the radial velocity measured with ALMA \citep{kospal2017c}. \label{fig:small}}
\end{figure}

Fig.~\ref{fig:small} shows the H$\alpha$ line and the median line profiles as a function of velocity for each species where we detected more than one line and where they were not affected by blends. The median H$_2$ line exhibits an approximately Gaussian shape that peaks at $-$15\,km\,s$^{-1}$ (close to the systemic velocity of $-7.8$\,km\,s$^{-1}$) and has a FWHM of 57.7\,km\,s$^{-1}$. We see no significant difference between the profiles of the $v$=1--0, 2--1, and 3--1 lines. The metallic lines have asymmetric profiles, with a steep rise at the blue side, a peak somewhere between $-$47 and $-$64\,km\,s$^{-1}$, a shoulder at about 0\,km\,s$^{-1}$, and a line wing that extends to above 100\,km\,s$^{-1}$ on the red side. The FWHM of the metallic lines is between 60 and 75\,km\,s$^{-1}$ for [FeII], [CaII], and [CrII], and between 103 and 113\,km\,s$^{-1}$ for [SII], [CI], [OI], and [NII]. The H$\alpha$ line has a profile similar to those of the forbidden metallic lines, but the component at about 0 velocity is even stronger, and the line is even wider (FWHM = 130\,km\,s$^{-1}$). The kind of complex line profile we observe in V346~Nor for the H$\alpha$ and the metallic lines can be understood as a combination of a high-velocity collimated jet component and a low-velocity wind component. 

As our XSHOOTER spectra are flux calibrated, we could measure the fluxes of the detected emission lines. For the symmetric H$_2$ lines, we fitted a Gaussian to each line and calculated the flux analytically from the amplitude and $\sigma$ of the fitted Gaussians. We estimated the uncertainty of the line flux from the residuals after subtracting the fitted Gaussian. For the asymmetric H$\alpha$ and metallic lines, we integrated the line fluxes between $-$150\,km\,s$^{-1}$ and +120\,km\,s$^{-1}$ and estimated the flux uncertainties by calculating the rms noise in the $-$600$\dots-$150\,km\,s$^{-1}$ and +120$\dots$+500\,km\,s$^{-1}$ ranges. The resulting line fluxes and uncertainties are presented in the Appendix in Tabs.~\ref{tab:h2} and \ref{tab:lines}.


\subsection{VISIR spectroscopy}

Fig.\,\ref{fig:visir} displays our $N$-band spectrum of V346\,Nor obtained with the VLT/VISIR instrument in 2016 February (Sect.\,\ref{sec:visirobs}). The spectrum exhibits a deep absorption feature, whose profile is reminiscent of the absorption feature measured toward Galactic Center sources and caused by submicron-sized amorphous silicate grains in the interstellar medium \citep{kemper2004}. From the profile we computed the optical depth at 9.7\,$\mu$m as ${\tau}_{9.7}$ = $-\log({F_{9.7}/F_{\rm cont}})$, where $F_{9.7}$ is the measured flux density at the canonical peak wavelength of the amorphous silicate feature, and $F_{\rm cont}$ is the estimated continuum level at 9.7\,$\mu$m. The continuum was determined by fitting a first order polynomial to the data points at 8\,$\mu$m and at ${\lambda}>12.5 \mu$m. The resulting optical depth value is ${\tau}_{9.7}$=0.82, which corresponds to an extinction of A$_{9.7}$=0.89\,mag, and A$_{V}$=16.4\,mag (adopting $A_V/A_{9.7}$=18.5, \citealt{mathis1990}). 

For comparison, we repeated the same computation for two Spitzer IRS low-resolution spectra measured on 2004 February 27 and 2006 April 16, and downloaded from the CASSIS public collection of processed and calibrated Spitzer spectra\footnote{https://cassis.sirtf.com/}. While these spectra also exhibit silicate absorption, the calculated optical depth values are significantly lower, ${\tau}_{9.7}$=0.36 and 0.44 for the earlier and later spectra, respectively. The deduced visual extinction values are A$_{V}$ = 7.3 and 8.8\,mag. These results are consistent with the simple reddened accretion disk models of \citet{kospal2017a}, which predicted that between 2003 and 2008 the brightness evolution of V346 Nor was governed by a parallel increase of the line-of-sight extinction (from 6.7 to 21.5\,mag) and a rise in the accretion rate (from 1$\times$10$^{-5}$ to 5$\times$10$^{-5}\,M_\odot$yr$^{-1}$). The relatively high visual extinction derived from our VISIR observation indicates that remarkable changes occurred in the line-of-sight extinction towards V346\,Nor, and that the extinction in the post-outburst phase was relatively high.

\section{Spectral analysis}
\label{sec:spectral}

\subsection{Spectral evolution in the post-outburst phase}
\label{sec:spectralevolution}

One of the earliest spectra of the (then) newly erupted V346~Nor was published by \citet{graham1985} using the CTIO 4\,m telescope and its Cassegrain spectrograph. They detected weak Li 671\,nm absorption, which we cannot check in our XSHOOTER spectrum, due to insufficient signal-to-noise ratio of the continuum at this wavelength. They also detected an H$\alpha$ line from V346~Nor with P~Cygni profile. The line had a broad, blueshifted absorption component, and a strong emission component near zero velocity. At this wavelength range we do not detect the continuum, therefore we cannot check the absorption component in our spectrum. However, our line profile must be different from theirs, because we have an emission peak at about $-$60\dots$-$50\,km\,s$^{-1}$ (same as the peak of the shock-related forbidden metallic lines), while \citet{graham1985} reported that the absorption component's red edge is at $-$60\,km\,s$^{-1}$. The H$\alpha$ line variability is not surprising, because already \citet{graham1985} noticed it by comparing several spectra taken over a year in 1983--84. However, we argue that the changes on longer time scales are much more substantial. \citet{graham1985} detected [OI], [NeII], and [SII] lines from HH~57, but not from V346~Nor, although they mentioned that the [SII] lines seem to extend also between the star and the HH object. We detected the [OI], [NeII], and [SII] lines at the stellar position. This suggests that V346~Nor did not display shock-related lines in 1983, but it did in 2015, indicating that shocked material close to the star must have formed some time between these two epochs.

\citet{reipurth1985} presented a low resolution optical spectrum taken in 1983 with the ESO\,3.6\,m telescope's Boller and Chivens spectrograph. The spectrum showed a red continuum with H$\alpha$ in absorption. The NaI doublet was clearly present in absorption, but they did not detect the Li line due to the low spectral resolution. We note that this spectrum does not show any emission lines, like the strong [OI] lines at 6300 and 6363$\AA$, the emission component of $H{\alpha}$, or the [SII] lines at 6716 and 6731$\AA$ that we detected. \citet{reipurth1985} also showed a low resolution CVF spectrum between 2--2.5$\,\mu$m obtained in 1983 with the ESO 1\,m telescope, and remarked on strong water vapor bands in the 2.0--2.2$\,\mu$m wavelength range. These bands are not apparent in our XSHOOTER spectrum.

A higher resolution $K$ band spectrum was published in \citet{reipurth1997}, taken in 1993 using NTT/IRSPEC. They tentatively detected Br$\gamma$ near 2.165$\,\mu$m, NaI near 2.206$\,\mu$m and 2.209$\,\mu$m, and FeI near 2.24$\,\mu$m, all in emission. We do not detect any of these lines, neither in emission nor in absorption in our XSHOOTER spectrum. \citet{reipurth1997} firmly detected CO bandhead emission, and speculated that it may arise from a neutral disk wind. We detect a faint CO bandhead absorption rather than emission, suggesting that the wind, if it existed in 1993, must have stopped by 2015. They detected the S(1), S(0), Q(1), and Q(3) lines of the 1--0 transition of H$_2$, all superimposed on the stellar continuum. We can confirm that these are indeed among the brightest of the H$_2$ lines we detect in the $K$ band, but we see many more lines. The H$_2$ lines did not become much stronger since 1993, but our XSHOOTER spectrum is more sensitive than the IRSPEC spectrum of \citet{reipurth1997}. It is possible that the shocks we observe towards V346~Nor already formed by 1993, although they were not present yet in 1983.

\citet{connelley2018} presented a detailed analysis of the near-infrared spectra of FUors. They introduced a new basis for classification based on whether or not an eruption was observed and on several spectroscopic criteria. Objects fulfilling these criteria are bona fide FUors. Objects that have spectra similar to bona fide FUors, but for which no eruption was observed are classified as FUor-like objects, and those that show some spectral similarities with bona fide FUors are called peculiar objects by \citet{connelley2018}. They observed V346~Nor on 2015 July 22, only a few days after our last XSHOOTER spectrum was taken. They detected no CO bandhead absorption and no water absorption bands (that should be present in bona fide FUors). On the other hand, they detected emission lines and weak metallic absorption lines (that should be absent in bona fide FUors). They could not check the presence of VO/TiO, Pa$\beta$ absorption, and He I absorption at 1.083$\,\mu$m, because their spectrum did not cover the necessary wavelengths. Our XSHOOTER spectrum allows us to re-evaluate these results. We confirm that water absorption is not present, but we do detect CO bandhead absorption. We checked the TiO and VO bands, but they are not present (should be present in bona fide FUors). We checked the Pa$\beta$ and HeI lines, and they are not present (both should be in absorption in bona fide FUors). \citet{connelley2018} claimed that they see emission lines and weak metallic absorption in V346~Nor. We cannot confirm the weak absorption lines, as we detected no absorption in our spectrum. Concerning the emission lines, we argue that most of them come from shocked material, and although we do not spatially resolve their origin, considering the very low transition probabilities, they must emerge from very tenuous material, and therefore cannot be very closely associated with the FUor itself. 





\subsection{Analysis of the CaII IRT lines}

\begin{figure}
\includegraphics[width=\columnwidth]{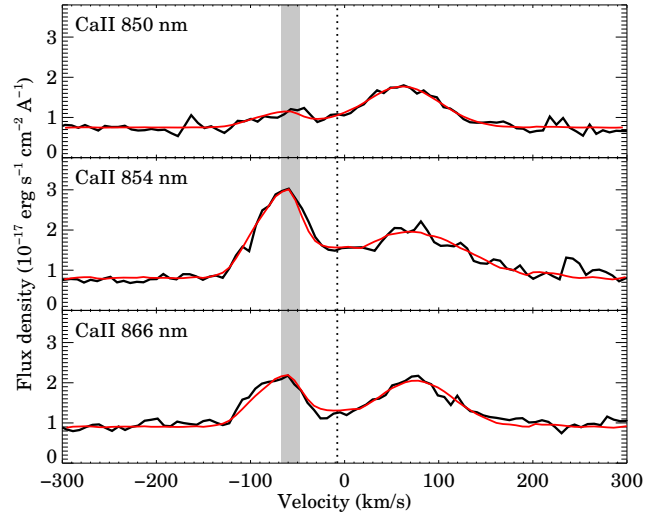}
\caption{Profiles of the three Ca IRT lines in V346~Nor. The gray band indicates the peak position of the shock-related forbidden metallic lines (cf.~Fig.~\ref{fig:small}), while the vertical dotted line marks the heliocentric systemic velocity of $-7.8$\,km\,s$^{-1}$.\label{fig:cairt}}
\end{figure}

As Fig.~\ref{fig:cairt} illustrates, the CaII IRT lines have two clearly distinct velocity components. The blueshifted one peaks at the same velocity as the shock-related forbidden metallic lines (such as the [CaII]) peak, marked by the gray stripe. Additionally, there is another, redshifted component that peaks around 70\,km\,s$^{-1}$. In order to determine the fluxes and peaks of these components, we fitted the sum of the scaled normalized [CaII] profile (from Fig.~\ref{fig:small}) and a Gaussian component to each line in the triplet. The resulting fit is plotted with red curves in Fig.~\ref{fig:cairt}, while the obtained line fluxes are presented in Tab.~\ref{tab:lines}. Interestingly, the ratio of the line peaks is 1:5.50:3.25 for the blueshifted (shock-like) component and 1:1.07:1.11 for the redshifted (Gaussian) component. This suggests that the shock-like component is more optically thin, closer to the theoretical optically thin limit of 1:8.847:4.884, while the Gaussian component is more optically thick, closer to the optically thick limit of 1:1:1. We calculated optical depth values for both components for each lines of the triplet, and found $\tau_{850\,\rm{nm}}$ = 0.19$\pm$0.05, $\tau_{854\,\rm{nm}}$ = 1.70$\pm$0.44, $\tau_{866\,\rm{nm}}$ = 0.94$\pm$0.24 for the shock component and $\tau_{850\,\rm{nm}}$ = 2.52$\pm$0.21, $\tau_{854\,\rm{nm}}$ = 22.3$\pm$1.8, $\tau_{866\,\rm{nm}}$ = 12.3$\pm$1.0 for the Gaussian component. This means that the Gaussian component is indeed about 13 times optically thicker than the shock component. 

\subsection{Analysis of the H$_2$ lines}

\begin{figure}[ht!]
\includegraphics[width=\columnwidth]{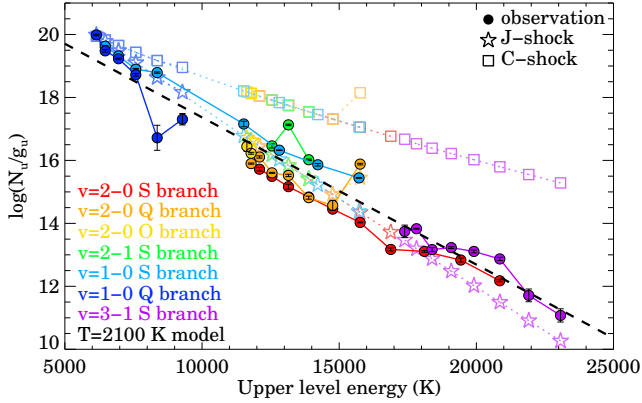}
\caption{Excitation diagram of H$_2$ emission from V346~Nor, fitted with a simple LTE model (dashed line). The observations are plotted with filled circles. As comparison, we plotted J-shock and C-shock models from \citet{smith1995} with open asterisks and squares, respectively, scaled to the $v=1-0$ Q(1) line.\label{fig:h2}}
\end{figure}

We detected a remarkable number of molecular hydrogen lines towards V346~Nor, with upper level energies up to 23\,000\,K. We used the measured line fluxes from Tab.~\ref{tab:h2}, together with the upper level energies and statistical factors to construct the excitation diagram of H$_2$. Fig.~\ref{fig:h2} shows this graph, using different colors for each branch. Our data points are fairly consistent with a straight line, expected for LTE. The inverse of the slope of the fitted line gives an LTE excitation temperature of $T_{\rm ex} = 2100 \pm 100\,$K. The measured temperature is similar to shocked regions associated with Class\,I/II YSOs \citep[e.g.,][]{takami2006,beck2008}. By comparing our line fluxes with those calculated by \citet{smith1995} for continuous (C-)shocks and jump (J-)shocks, our results are essentially consistent with a C-shock, as expected for dense clouds (Fig.~\ref{fig:h2}). Interestingly, the H$_2$ lines in HH~57 are more consistent with a J-shock \citep{eisloffel2000}. This is consistent with the two-shock picture of working surfaces \citep{hollenbach1997}.

\subsection{Analysis of the [FeII] lines}

\begin{figure*}
\includegraphics[width=0.33\textwidth]{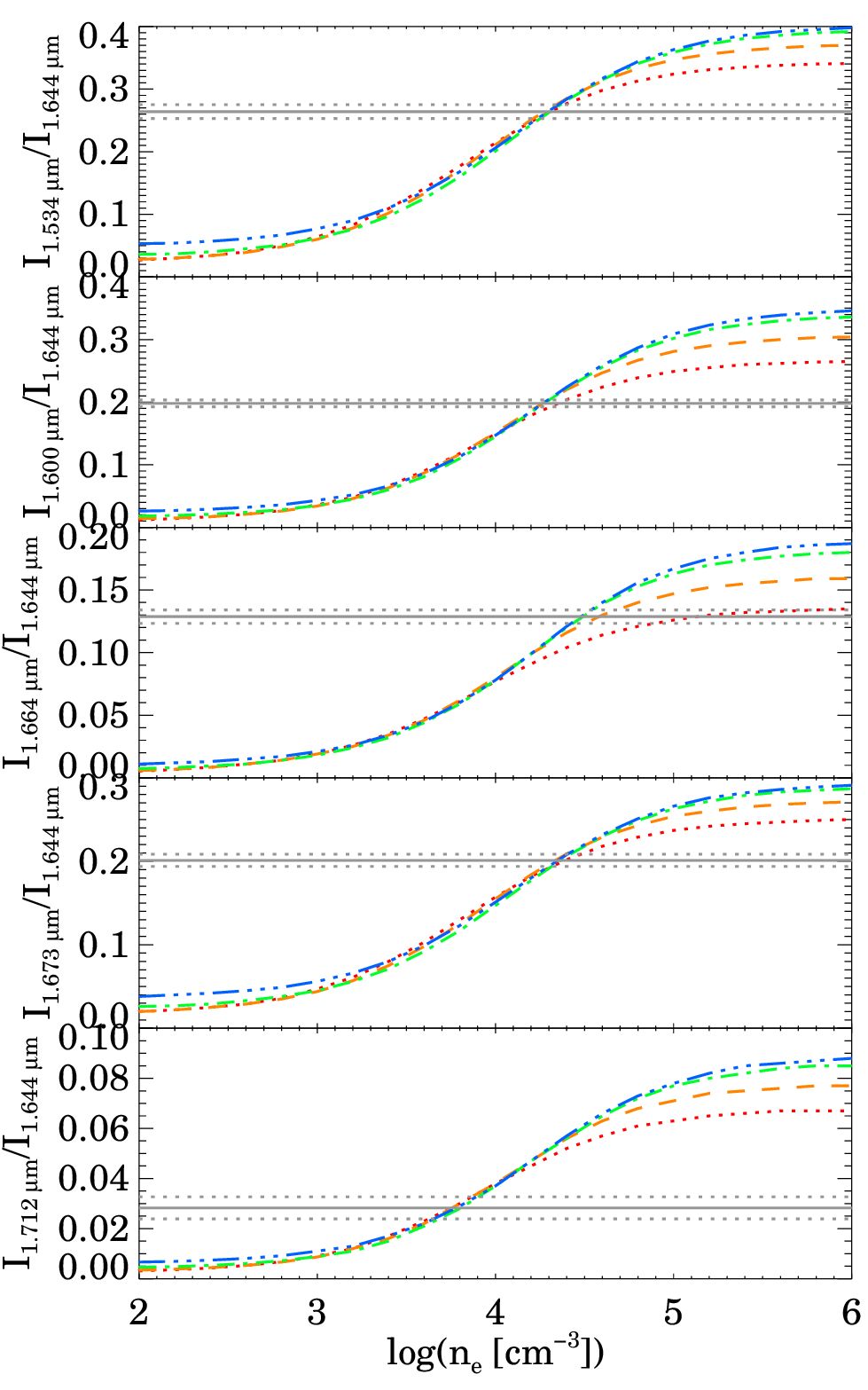}
\includegraphics[width=0.33\textwidth]{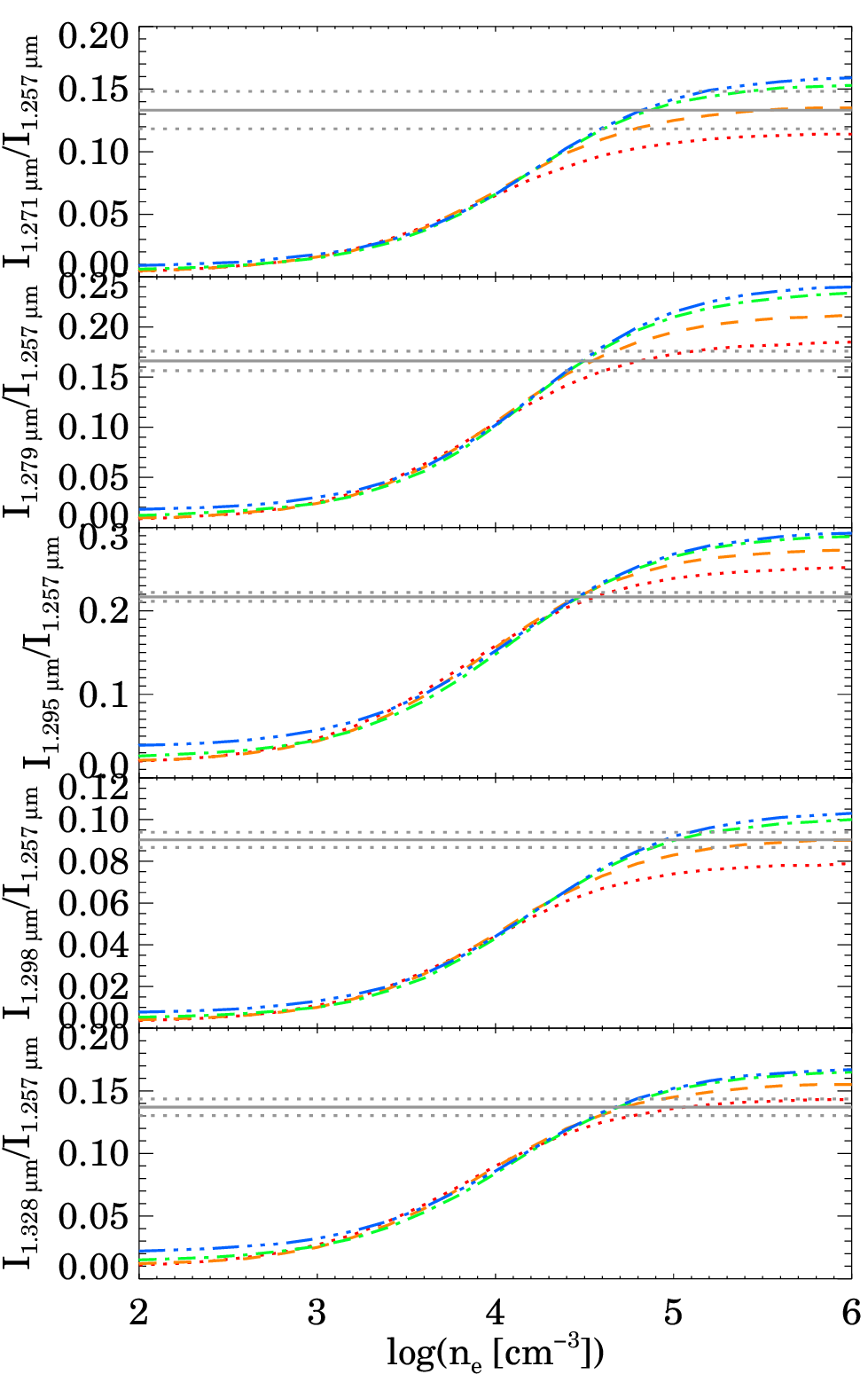}
\includegraphics[width=0.33\textwidth]{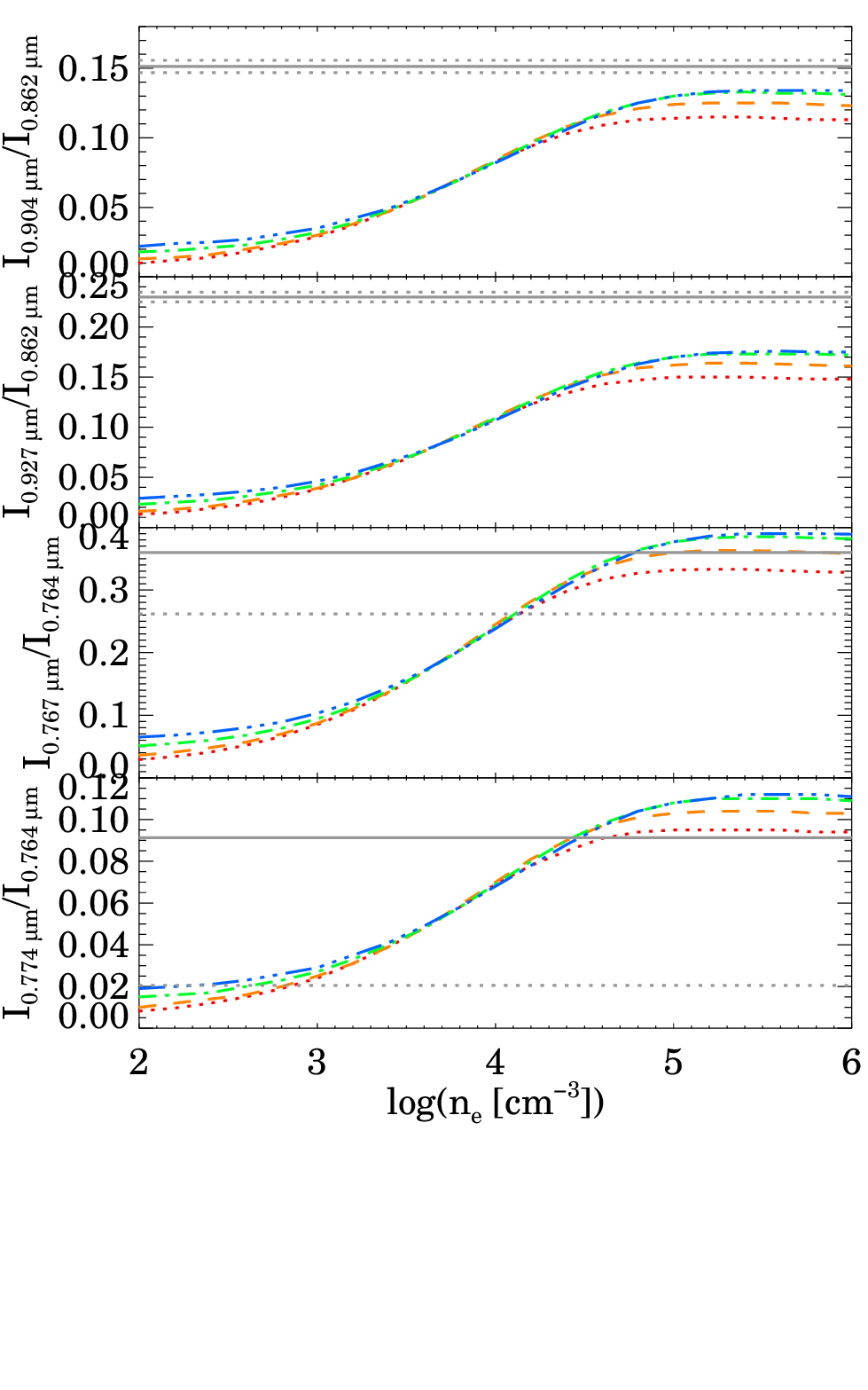}
\caption{Line ratios for different [FeII] lines observed for V346~Nor in the $H$ band (left), in the $J$ band (middle) and in the visual band (right). The dotted red, dashed orange, dash-dotted green, and dash-triple dotted blue curves are models for $T_{\rm e}$ = 3000, 5000, 10\,000, and 20\,000\,K, respectively. The gray horizontal lines mark the observed line ratios, with gray dotted lines showing the $\pm1\sigma$ uncertainties. \label{fig:fe}}
\end{figure*}

We used the CHIANTI Database V9.0\footnote{http://chiantidatabase.org/} to calculate model line ratios for our observed lines for different $n_{\rm e}$ electron densities and $T_{\rm e}$ electron temperatures \citep{dere1997,dere2019}. Fig.~\ref{fig:fe} shows the model curves as a function of $N_{\rm e}$ for different temperatures, together with the observed line ratios. By looking at Tab.~\ref{tab:lines}, it is evident that the [FeII] lines detected can be grouped into transitions between the following configurations and terms:\\
\begin{itemize}
    \item 3d$^7$ a $^4$F and 3d$^7$ a $^2$G (between 716 and 745\,nm)
    \item 3d$^6$($^5$D)4s a $^6$D and 3d$^7$ a $^4$P (between 763 and 773\,nm)
    \item 3d$^7$ a $^4$F and 3d$^7$ a $^4$P (between 862 and 947\,nm)
    \item 3d$^6$($^5$D)4s a $^6$D and 3d$^6$($^5$D)4s a $^4$D (between 1257 and 1328\,nm)
    \item 3d$^7$ a $^4$F and 3d$^6$($^5$D)4s (between 1533 and 1711\,nm)
\end{itemize}
For each group we took the strongest line and calculated the model and observed flux ratios for all the other lines in the group. Some of these are illustrated in Fig.~\ref{fig:fe}. Line ratios in the $H$ band (left panel) suggest an electron density of 10$^4\dots$10$^{4.5}$\,cm$^{-3}$, while line ratios in the $J$ band (middle panel) suggest electron density closer to 10$^5$\,cm$^{-3}$. The observations are more consistent with higher temperature models (10\,000\dots20\,000\,K) rather than lower temperature models (3000\dots5000\,K). The flux ratios of the optical lines are  inconsistent with the model predictions.

\subsection{Analysis of the CO lines}

\begin{figure}
\includegraphics[height=\columnwidth,angle=90]{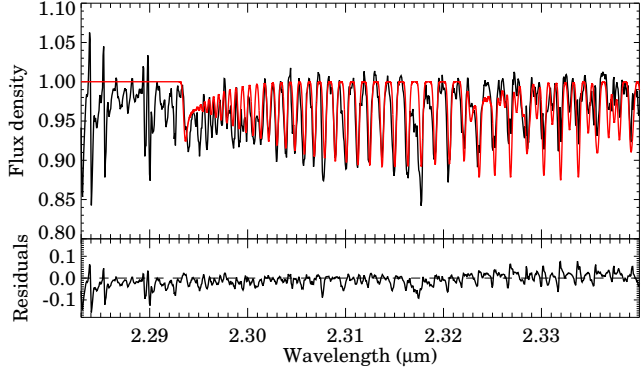}
\caption{Part of our XSHOOTER $K$-band spectrum of V346~Nor showing cool overtone CO absorption (black). The model curve (red) is an LTE slab model with $T$=1700\,K. \label{fig:co}}
\end{figure}

FUors typically show the overtone CO feature around 2.3$\,\mu$m in absorption, while several T~Tauri stars show this characteristic feature in emission. At low spectral resolution, the feature appears as a blend of many individual rovibrational lines corresponding to the v=2--0, 3--1, 4--2, 5--3, ... and ${\Delta}J = {\pm}1$ transitions. Where the density and temperature is high enough for CO to be collisionally excited ($n_{\rm H} > 10^{10}$\,cm$^{-3}$, $T \gtrsim 2000$\,K, \citealt{scoville1980}), the high-$J$ transitions are highly excited, and blend together to form the usual ``bandhead''. Although no strong feature is visible, there are unmistakable lines in the 2.29--2.34$\,\mu$m wavelength range. This seems to suggest that in V346~Nor, mainly the low-$J$ transitions are excited (Fig.~\ref{fig:co}). While in this case the LTE approach may not be entirely appropriate, to confirm the detection of CO, we compared the observed spectrum with a simple isothermal slab model and found a good match using $T = 1700$\,K and $N_{\rm CO} = 7{\times}10^{19}$\,cm$^{-2}$. After normalizing our spectrum with a third order Legendre polynomial and shifting it by 2.9\,km\,s$^{-1}$, our model fits the observed spectrum fairly well, suggesting absorption by cool ($\sim 1700$\,K) CO in the V346~Nor system. The small shift means that the velocity of the CO emitting gas is consistent with the systemic velocity within the uncertainties of the XSHOOTER wavelength calibration. Our results imply that there cannot be hotter CO in the system, because then the strong CO bandhead emission would overwhelm the faint low-$J$ absorption lines we observed. 

\section{Discussion}
\label{sec:disc}

\subsection{The post-outburst evolution of V346~Nor}
\label{sec:lightcurveevolution}

In this section we synthesize the results obtained in the previous analyses, and attempt to draw a picture of the post-outburst evolution of V346 Nor following its brightness minimum in 2010-11. Mainly based on the light curves in Fig.~\ref{fig:lightcurve} we divide this period into six parts.

\paragraph{Period 1: first re-brightening from 2010 to mid-2013} This period was also briefly described in \citet{kraus2016} and \citet{kospal2017a}. The brightness minimum lasted until late 2011, then a gradual brightening started. During the next $\sim$1.5 year the source's light increased by $\sim$1.8\,mag in the $K_{\rm s}$ band. When the source entered the minimum, it faded by $\sim$4.4\,mag between 2008 April and 2010 March (23 months) in the $K_{\rm s}$ band. The average fading rate was ${\Delta}K$/${\Delta}t$ = 0.19\,mag/month. During the re-brightening it was ${\Delta}K$/${\Delta}t$ = $-$0.1\,mag/month, slightly slower, but still comparable to the decay rate. The source seem to have brightened also at the WISE wavelengths, however, there is a known artificial offset between the ALLWISE (2010) and the NEOWISE (2014--) magnitudes for bright stars \citep{cotten2016}, and it might affect V346~Nor, especially in the W2 band. Thus, the WISE flux change should be considered with care. From the near-infrared color-magnitude in Fig.~\ref{fig:tcd} (a), it seems that the source moved back from the 2010 minimum toward the 2008 position, suggesting that the same process which brought the source to the minimum operates here, but reversed. The similarity of the fading and brightening rates supports this speculation. The color changes exclude decreasing line-of-sight extinction as the reason. According to the modeling in \citet{kospal2017a}, when V346~Nor entered the minimum, the accretion rate dropped while the extinction remained unchanged and high, $A_V \sim 21.5$\,mag. If a reversed process was in operation in 2012--13, then we propose that the accretion rate gradually increased while the line-of-sight extinction was relatively constant and still high.


\paragraph{Period 2: the plateau between mid-2013 and mid-2016} The brightening of V346~Nor temporarily stopped in mid-2013. Both the $I$ and the $K_{\rm s}$ light curves indicate that the source was approximately constant (within $\sim$0.3\,mag) from mid-2013 until mid-2015. In Sec.~\ref{sec:colorevolution} we computed representative average $IJHKW1W2$ magnitudes for this plateau phase and plotted the SED in Fig.~\ref{fig:sed}. In comparison with the outburst SED, the difference is highest in the $I$ band and smallest in the $L$/$W1$ band, showing a exponentially decreasing trend with wavelength. After correcting this difference SED for $A_V$=21.5\,mag extinction, the result resembles a hot accretion disk, supporting our view that the first brightening was an accretion event. Another proof for active accretion is our XSHOOTER spectrum, which did not display any photospheric absorption features due to strong veiling. The bandhead emission from hot CO that was visible in 1993 disappeared by 2015, therefore the wind component theorized by \citet{reipurth1997} must have stopped. Our VISIR spectrum from early 2016 suggests that the source is embedded. The derived $A_V$=16.4\,mag is comparable to the $A_V$=21.5\,mag computed from our accretion disk models for the minimum in 2010 \citep{kospal2017a}.

\paragraph{Period 3: V346 Nor becomes active between mid-2016 and mid-2017} In this period V346~Nor becomes more variable than previously, although no general brightening trend can be seen, expect maybe in the $I$ band. In the first half of 2017  a localized brightness peak can be seen on top of a linear baseline. The slope of the baseline changes with wavelength: rising at shorter wavelengths and decreasing at longer wavelength. The peak was approximately gray: its amplitude is $\sim$ 0.5\,mag in $JHK_{\rm s}$, suggesting accretion fluctuations on a few months time scale.


\paragraph{Period 4: the second brightening between mid-2017 and mid-2018} In August 2017, V346~Nor suddenly started a general brightening  at all wavelengths. The magnitude changes at different wavelengths again indicate gray brightening, which excludes extinction variations, and point to gradually increasing accretion rate instead. The brightening rate in the $K_{\rm s}$ band was  ${\Delta}K_{\rm s}/{\Delta}t\sim0.08$\,mag/month, very close to the rate measured during the first brightening in 2012--13 (0.1\,mag/month). This suggests that the same physical mechanism was in operation in both brightening periods. We note that by the end of this period the source was still several magnitudes fainter than the outburst level in 1998--2004.


\paragraph{Period 5: a rapid fading in the second half of 2018} Between 2018 July and October, V346~Nor unexpectedly faded at all wavelengths. The amplitude of the fading was higher at shorter wavelengths, meaning that the source became redder when fainter. The color trends of this variability were close to, but not entirely consistent with the reddening path. A tentative interpretation of this phenomenon could be an eclipsing event, where an obscuring dust cloud causes the fading of the system. In this picture, the deviation from the reddening path could be due to the increasing contribution from scattered light at shorter wavelengths, similarly to the blueing effect observed in UXor-type variables \citep[cf.][]{natta2000}.

\paragraph{Period 6: the state of V346~Nor in 2019} By early 2019 the sudden fading observed at the end of 2018 stopped and V346~Nor appeared brighter again. This is followed by a very peculiar behavior: in the $H$ and $K_{\rm s}$ bands the source has been brightening, while in the $I$ and $J$ bands, it has been fading. We have then two possible explanations for this phenomenon (1) the accretion is still increasing, as suggested by the $HK$ light curves, but at shorter wavelengths some other effect balances it, perhaps increasing extinction by a small dust cloud; (2) the increase of accretion rate stopped in 2019, and the continuing flux increase in the $HK$ bands is related to the appearance of some warm dust component that emits mainly longer wavelengths. In conclusion, V346~Nor seem to exhibit increasingly complex flux changes in the optical-infrared regime.


\subsection{A new jet emanating from V346~Nor?}

We detected several shock-related forbidden metallic and H$_2$ lines in our XSHOOTER spectrum of V346~Nor towards the stellar position. We note that although HH~57 was also included in the slit in some of our spectra, we will analyze those data in a later paper (Kraus et al.~in prep.), while here we focus on the line of sight towards V346~Nor only. The comparison of our spectrum with earlier data reveals that these lines were not present towards V346~Nor in 1984-85 (Sec.~\ref{sec:spectralevolution}, see also \citealt{graham1985}, \citealt{reipurth1985}), while they were clearly detected in 1993 \citep{reipurth1997}, with similar brightness as in our spectrum from 2015.

In our spectra, the forbidden emission comes from within 1-2$''$ of the stellar position, meaning that there should be shocked material in the $\sim$1000\,au vicinity of V346~Nor. The various lines indicate multiple components: a hotter part whose bulk moves with a line of sight velocity of about $-$60\,km\,s$^{-1}$ towards us and emits forbidden metallic lines, and a cooler part, that is more or less stationary compared to the young star and emits molecular hydrogen lines. A similar velocity difference between [FeII] and H$_2$ emission was observed in several Class\,I objects by \citet{davis2003,takami2006}. The usual interpretation is that the [FeII] lines originates from Herbig--Haro-type shocked material along the jet axis, while the H$_2$ emission comes from closer to the star, at the interface of the jet and the dense circumstellar material. It is tempting to associate the detected outflowing material with a new jet emanating from V346~Nor due to the enhanced accretion during its 1980--2010 outburst. We cannot explain why the shocks did not appear immediately after the beginning of the outburst, but only with a certain time delay. A possible explanation could be that the shocks did not appear until the outflowing material hit denser matter in, e.g., the outflow cavity walls. Indeed, CO molecular line observations with ALMA showed ellipse-shaped outflow cavities around V346~Nor with opening angles between 40$^{\circ}$ and 80$^{\circ}$ \citep{kospal2017c}. 

\subsection{Recurrent outbursts in FUors and related objects}

V346~Nor is not the only young eruptive object that showed a temporary halt in the accretion rate. Apart from the clearly repetitive outbursts of EX Lupi-type objects (EXors), there are other young eruptive stars that showed multiple outbursts. V1647~Ori is an embedded object that was in outburst between 2004 and 2006, then again between 2008 and 2018, and photographic plates suggest an outburst in 1966 as well \citep{aspin2006, acosta2007, ninan2013, giannini2018}. It is not a bona fide FUor, because its light curves and spectra show a mix of characteristics from both FUors and EXors. V1647~Ori drives both a molecular outflow with a rather narrow (30$^{\circ}$) opening angle \citep{principe2018} and is associated with the Herbig-Haro object HH~23 \citep{eisloffel1997}. 

V899~Mon is a recently discovered young outbursting star \citep{wils2009}, whose photometric and spectroscopic characteristics also fall between classical FUors and EXors \citep{ninan2015}. V899~Mon displayed a slow brightening until 2010, when its $R$ magnitude suddenly dropped by 4\,mag from 12 to almost 16\,mag for about two years, after which its brightness stabilised at about $R=13$\,mag. \citet{ninan2015} interpret this as a short-duration halt in the accretion. High resolution optical spectroscopy of V899~Mon revealed forbidden emission lines, such as the [OI] lines at 6300 and 6363\,${\AA}$, and the [FeII] line at 7155$\,{\AA}$. We detected these lines in V346~Nor as well. In V899~Mon, the  profiles of the forbidden emission lines indicate outflowing material with velocities up to $-$500\,km\,s$^{-1}$ \citep{ninan2015}. Our CO observations obtained with APEX and ALMA suggest that V899~Mon also drives a powerful molecular outflow (K\'osp\'al et al.~in prep.).

It appears that this group of objects share some common photometric and spectroscopic characteristics with some similarities in their accretion and outflow properties. Their further study may reveal important information about young eruptive stars that so far remained elusive. Currently ongoing and future sky surveys like WISE, Gaia, TESS, and LSST will provide light curves of many young stars, allowing us to discover not only new eruptions, but fadings that may signal the end or temporary halt in the eruption of a young star. Their further spectrosopic follow-up may provide new insights into the post-FUor state, which may reveal whether our results on V346~Nor could be applied to further objects. A significant diversity may make it necessary to invoke more than one outburst mechanisms.

\section{Conclusions}
\label{sec:conclusions}

In this paper, we reported on a multi-epoch and multi-wavelength study of V346~Nor after its deep minimum in 2010--11. During this post-outburst phase, we found that the evolution of the source was governed by four important physical processes: variations in the accretion rate, the launch of a new jet, variations in the line-of-sight extinction, and the appearance of new warm material in the system. Following the minimum, the accretion rate in V346~Nor has increased. It is difficult to quantify how much this increase was, because there is a degeneracy between the accretion rate and the line-of-sight extinction. The accretion is probably significantly higher than it was in the minimum (because of the significant veiling), but less than during the outburst (because the source is still fainter). The increase of accretion was not steady: shorter rising periods were interrupted by years of constant brightness. At some point between the beginning of the outburst (around 1980) and 1993, a new jet was launched, as evidenced by several strong forbidden emission lines of various metals and molecular hydrogen. This will probably propagate further from the star and may appear as a new HH-object in a few hundred to few thousand years from now. The mid-infrared spectra indicate that in post-outburst, the line-of-sight extinction is higher than in outburst. According to our accretion disk modeling \citep{kospal2017a}, the accretion rate increase probably started already around 2008, before the minimum, and it is still high. Therefore, the post-outburst phase seems to differ from the outburst in that there is more material close to the center of the system. It is an open question if the extinction variations are related to the physical mechanism causing the minimum. The significantly increased $K_{\rm s}$-band brightness in 2019 and the high continuum level of our VISIR spectrum already in 2016 suggest the appearance of warm dusty ($T<1500$\,K) material in the system. We speculate that this may be a refilling process that allows the continuation of the outburst in the future. 
It is of course an open question how the post-outburst evolution of V346~Nor will continue. Our latest measurements do not reveal unambiguously if it will continue to brighten to a second outburst or will fade back to quiescence.



\acknowledgments

This project has received funding from the European Research Council (ERC) under the European Union's Horizon 2020 research and innovation programme under grant agreement No 716155 (SACCRED) and No 639889 (ImagePlanetFormDiscs). This project is based on observations collected at the European Southern Observatory under ESO programme 095.C-0765. CHIANTI is a collaborative project involving George Mason University, the University of Michigan (USA), University of Cambridge (UK) and NASA Goddard Space Flight Center (USA). We gratefully acknowledge the help of Dr.~Valentin D.~Ivanov (ESO) in helping us with the XSHOOTER data reduction. The OGLE project has received funding from the National Science
Centre, Poland, grant MAESTRO 2014/14/A/ST9/00121 to A.U.

\facilities{VLT:Kueyen, VLT:Melipal, CTIO:1.3m, ESO:VISTA, WISE}

\software{ESO XSHOOTER pipeline, ESO VISIR pipeline, EsoReflex, EsoRex, molecfit \citep{smette2015, kausch2015}, CHIANTI \citep{dere1997, dere2019}}

\appendix

\section{Photometry of V346~Nor}

Tab.~\ref{tab:phot} contains our optical and near-infrared photometry of V346~Nor.

\startlongtable
\begin{deluxetable*}{ccccccc}
\tablecaption{Optical and near-IR photometry of V346~Nor.\label{tab:phot}}
\tablehead{
\colhead{Date} & \colhead{JD $-$ 2450000.5} & \colhead{$I$} &
\colhead{$J$} & \colhead{$H$} & \colhead{$K_{\rm s}$} & \colhead{Instrument}}
\startdata
2013-05-25 & 6438.27 & $20.60\pm0.46$&\dots&\dots&\dots& OGLE \\    
2013-06-23 & 6467.27 & $19.88\pm0.30$&\dots&\dots&\dots& OGLE \\
2013-08-16 & 6521.08 & $20.74\pm0.41$&\dots&\dots&\dots& OGLE \\
2014-05-15 & 6793.28 & $20.92\pm0.55$&\dots&\dots&\dots& OGLE \\
2014-05-16 & 6794.29 & $19.95\pm0.28$&\dots&\dots&\dots& OGLE \\
2014-05-17 & 6795.26 & $20.11\pm0.17$&\dots&\dots&\dots& OGLE \\
2014-09-30 & 6931.03 & $20.28\pm0.34$&\dots&\dots&\dots& OGLE \\
2015-02-08 & 7062.35 & $20.31\pm0.25$&\dots&\dots&\dots& OGLE \\
2015-02-11 & 7065.32 & $20.45\pm0.27$&\dots&\dots&\dots& OGLE \\
2015-02-13 & 7067.33 & $20.42\pm0.32$&\dots&\dots&\dots& OGLE \\
2015-02-14 & 7068.35 & $20.41\pm0.23$&\dots&\dots&\dots& OGLE \\
2015-02-16 & 7070.32 & $20.59\pm0.24$&\dots&\dots&\dots& OGLE \\
2015-02-20 & 7074.32 & $20.12\pm0.29$&\dots&\dots&\dots& OGLE \\
2015-02-23 & 7077.31 & $20.27\pm0.23$&\dots&\dots&\dots& OGLE \\
2015-02-24 & 7078.31 & $20.51\pm0.26$&\dots&\dots&\dots& OGLE \\
2015-02-28 & 7082.28 & $20.55\pm0.34$&\dots&\dots&\dots& OGLE \\
2015-03-01 & 7083.29 & $20.50\pm0.31$&\dots&\dots&\dots& OGLE \\
2015-03-03 & 7085.25 & $20.66\pm0.39$&\dots&\dots&\dots& OGLE \\
2015-03-03 & 7088.26 & $20.32\pm0.24$&\dots&\dots&\dots& OGLE \\
2015-03-07 & 7089.27 & $20.26\pm0.24$&\dots&\dots&\dots& OGLE \\
2015-03-09 & 7091.26 & $20.94\pm0.48$&\dots&\dots&\dots& OGLE \\
2015-03-10 & 7092.26 & $20.13\pm0.20$&\dots&\dots&\dots& OGLE \\
2015-03-11 & 7093.25 & $20.52\pm0.29$&\dots&\dots&\dots& OGLE \\
2015-03-15 & 7097.25 & $20.77\pm0.30$&\dots&\dots&\dots& OGLE \\
2015-03-16 & 7098.25 & $20.57\pm0.26$&\dots&\dots&\dots& OGLE \\
2015-03-18 & 7100.24 & $20.43\pm0.27$&\dots&\dots&\dots& OGLE \\
2015-03-26 & 7108.23 & $20.12\pm0.24$&\dots&\dots&\dots& OGLE \\
2015-03-28 & 7110.22 & $20.08\pm0.15$&\dots&\dots&\dots& OGLE \\
2015-03-31 & 7112.72 &\dots&\dots&\dots&$10.79\pm0.13$&  VISTA \\
2015-03-31 & 7113.19 & $20.44\pm0.30$&\dots&\dots&\dots& OGLE \\
2015-04-02 & 7115.20 & $20.14\pm0.25$&\dots&\dots&\dots& OGLE \\
2015-04-06 & 7119.19 & $20.34\pm0.35$&\dots&\dots&\dots& OGLE \\
2015-04-12 & 7125.18 & $19.87\pm0.16$&\dots&\dots&\dots& OGLE \\
2015-04-14 & 7127.17 & $20.53\pm0.28$&\dots&\dots&\dots& OGLE \\
2015-04-17 & 7130.15 & $20.16\pm0.18$&\dots&\dots&\dots& OGLE \\
2015-04-19 & 7132.14 & $19.96\pm0.20$&\dots&\dots&\dots& OGLE \\
2015-04-21 & 7134.12 & $20.47\pm0.37$&\dots&\dots&\dots& OGLE \\
2015-04-22 & 7135.14 & $20.65\pm0.32$&\dots&\dots&\dots& OGLE \\
2015-04-23 & 7136.13 & $20.24\pm0.18$&\dots&\dots&\dots& OGLE \\
2015-04-24 & 7137.13 & $20.26\pm0.18$&\dots&\dots&\dots& OGLE \\
2015-04-25 & 7138.11 & $20.22\pm0.32$&\dots&\dots&\dots& OGLE \\
2015-04-27 & 7140.11 & $19.59\pm0.15$&\dots&\dots&\dots& OGLE \\
2015-04-28 & 7141.10 & $20.18\pm0.22$&\dots&\dots&\dots& OGLE \\
2015-04-29 & 7142.13 & $19.85\pm0.17$&\dots&\dots&\dots& OGLE \\
2015-04-30 & 7143.11 & $20.26\pm0.30$&\dots&\dots&\dots& OGLE \\
2015-05-03 & 7146.10 & $20.19\pm0.35$&\dots&\dots&\dots& OGLE \\
2015-05-05 & 7148.24 & $20.20\pm0.26$&\dots&\dots&\dots& OGLE \\
2015-05-08 & 7151.35 & $20.46\pm0.39$&\dots&\dots&\dots& OGLE \\
2015-05-10 & 7153.08 & $20.91\pm0.38$&\dots&\dots&\dots& OGLE \\
2015-05-12 & 7155.09 & $20.01\pm0.20$&\dots&\dots&\dots& OGLE \\
2015-05-13 & 7156.07 & $20.29\pm0.23$&\dots&\dots&\dots& OGLE \\
2015-05-14 & 7157.08 & $20.42\pm0.26$&\dots&\dots&\dots& OGLE \\
2015-05-15 & 7158.06 & $20.83\pm0.41$&\dots&\dots&\dots& OGLE \\
2015-05-16 & 7159.07 & $20.21\pm0.24$&\dots&\dots&\dots& OGLE \\
2015-05-17 & 7160.05 & $20.18\pm0.29$&\dots&\dots&\dots& OGLE \\
2015-05-19 & 7162.06 & $20.70\pm0.33$&\dots&\dots&\dots& OGLE \\
2015-05-20 & 7163.05 & $20.38\pm0.38$&\dots&\dots&\dots& OGLE \\
2015-05-21 & 7164.09 & $20.00\pm0.18$&\dots&\dots&\dots& OGLE \\
2015-05-23 & 7166.05 & $20.44\pm0.37$&\dots&\dots&\dots& OGLE \\
2015-05-27 & 7170.02 & $20.38\pm0.32$&\dots&\dots&\dots& OGLE \\
2015-05-30 & 7173.02 & $20.32\pm0.41$&\dots&\dots&\dots& OGLE \\
2015-05-31 & 7174.02 & $19.66\pm0.27$&\dots&\dots&\dots& OGLE \\
2015-06-05 & 7179.02 & $20.43\pm0.30$&\dots&\dots&\dots& OGLE \\
2015-06-08 & 7182.01 & $20.64\pm0.34$&\dots&\dots&\dots& OGLE \\
2015-06-09 & 7183.05 & $20.03\pm0.21$&\dots&\dots&\dots& OGLE \\
2015-06-09 & 7183.05 & $19.97\pm0.25$&\dots&\dots&\dots& OGLE \\
2015-06-12 & 7186.01 & $19.88\pm0.27$&\dots&\dots&\dots& OGLE \\
2015-06-13 & 7187.00 & $20.63\pm0.43$&\dots&\dots&\dots& OGLE \\
2015-06-15 & 7189.00 & $20.79\pm0.42$&\dots&\dots&\dots& OGLE \\
2015-06-17 & 7190.99 & $20.66\pm0.49$&\dots&\dots&\dots& OGLE \\
2015-06-20 & 7193.98 & $20.33\pm0.18$&\dots&\dots&\dots& OGLE \\
2015-07-20 & 7223.68 &\dots&\dots&\dots&$10.33\pm0.01$&  VISTA \\
2016-02-06 & 7425.32 & $20.94\pm0.45$&\dots&\dots&\dots& OGLE \\
2016-02-12 & 7431.31 & $20.72\pm0.35$&\dots&\dots&\dots& OGLE \\
2016-02-15 & 7434.32 & $20.10\pm0.22$&\dots&\dots&\dots& OGLE \\
2016-02-19 & 7438.30 & $20.45\pm0.27$&\dots&\dots&\dots& OGLE \\
2016-02-21 & 7440.30 & $20.19\pm0.24$&\dots&\dots&\dots& OGLE \\
2016-02-24 & 7443.31 & $20.49\pm0.33$&\dots&\dots&\dots& OGLE \\
2016-02-27 & 7446.29 & $20.27\pm0.21$&\dots&\dots&\dots& OGLE \\
2016-03-01 & 7449.28 & $20.07\pm0.30$&\dots&\dots&\dots& OGLE \\
2016-03-02 & 7450.23 & $20.03\pm0.19$&\dots&\dots&\dots& OGLE \\
2016-03-10 & 7458.26 & $20.10\pm0.19$&\dots&\dots&\dots& OGLE \\
2016-03-13 & 7461.26 & $20.76\pm0.27$&\dots&\dots&\dots& OGLE \\
2016-03-18 & 7466.25 & $19.93\pm0.25$&\dots&\dots&\dots& OGLE \\
2016-03-22 & 7470.22 & $19.93\pm0.28$&\dots&\dots&\dots& OGLE \\
2016-03-31 & 7479.27 & $20.06\pm0.21$&\dots&\dots&\dots& OGLE \\
2016-06-06 & 7546.06 & $>19.59$ & $15.07\pm0.03$ & $12.67\pm0.01 $ & $10.22\pm0.02$ & SMARTS$^a$\\ 
2016-06-20 & 7560.22 & $19.98\pm0.26$&\dots&\dots&\dots& OGLE \\
2016-06-28 & 7568.04 & $19.90\pm0.20$&\dots&\dots&\dots& OGLE \\
2016-07-10 & 7579.51 &\dots&\dots&\dots&$9.94\pm0.01$&  VISTA \\
2016-07-13 & 7582.65 &\dots&\dots&\dots&$9.82\pm0.01$&  VISTA \\
2016-08-09 & 7610.00 & $>19.27$ & $14.44\pm0.02$ & $12.17\pm0.01$ & $9.75\pm0.02$ & SMARTS$^a$\\
2016-08-22 & 7622.00 & \dots&\dots&\dots&$9.90\pm0.02$& IRIS \\
2016-08-25 & 7625.00 & \dots &$14.52\pm0.02$ & $12.26\pm0.02$ & $9.92\pm0.02$ & IRIS \\
2017-02-01 & 7786.25 & $>19.62$ & $15.17\pm0.02$ & $12.81\pm0.02$ & $10.11\pm0.02$& SMARTS\\
2017-02-10 & 7795.32 & $19.60\pm0.20$&\dots&\dots&\dots& OGLE \\
2017-02-18 & 7803.31 & $20.08\pm0.18$&\dots&\dots&\dots& OGLE \\
2017-03-02 & 7815.25 & $>19.58$ & $15.07\pm0.03$  & $12.76\pm 0.01$ & $10.11\pm0.02$ & SMARTS\\
2017-03-11 & 7824.27 & $20.02\pm0.21$&\dots&\dots&\dots& OGLE \\
2017-03-25 & 7838.20 & $19.80\pm0.16$&\dots&\dots&\dots& OGLE \\
2017-04-02 & 7846.23 & $19.56\pm0.12$&\dots&\dots&\dots& OGLE \\
2017-04-02 & 7846.25 & $>19.77$ & $15.03\pm0.02$ & $12.77\pm0.01$ & $10.14\pm0.02$ & SMARTS\\
2017-05-02 & 7876.25 & $>19.75$ &  $14.95\pm0.02$ & $12.61\pm0.01$ & $10.21\pm0.03$ & SMARTS\\ 
2017-06-03 & 7908.00 & $>19.16$ & $14.65\pm0.02$ & $12.38\pm0.02$ & $9.94\pm0.02$ & SMARTS\\
2017-06-04 & 7909.02 & $19.16\pm0.17$&\dots&\dots&\dots& OGLE \\
2017-06-14 & 7918.64 &\dots&\dots&\dots&$10.08\pm0.01$&  VISTA \\
2017-06-17 & 7922.05 &\dots&\dots&\dots&$9.96\pm0.01$&  VISTA \\
2017-06-20 & 7925.49 &\dots&\dots&\dots&$10.18\pm0.01$&  VISTA \\
2017-06-21 & 7926.57 &\dots&\dots&\dots&$10.24\pm0.01$&  VISTA \\
2017-06-28 & 7933.64 &\dots&\dots&\dots&$10.04\pm0.01$&  VISTA \\
2017-06-30 & 7935.69 &\dots&\dots&\dots&$10.17\pm0.01$&  VISTA \\
2017-07-02 & 7937.65 &\dots&\dots&\dots&$10.30\pm0.01$&  VISTA \\
2017-07-05 & 7940.63 &\dots&\dots&\dots&$10.39\pm0.01$&  VISTA \\
2017-07-07 & 7942.00 & $>19.39$ & $14.88\pm0.02$ & $12.68\pm0.01$ & $10.28\pm0.02$ & SMARTS\\
2017-07-07 & 7942.65 &\dots&\dots&\dots&$10.15\pm0.01$&  VISTA \\
2017-07-14 & 7948.66 &\dots&\dots&\dots&$10.16\pm0.01$&  VISTA \\
2017-08-05 & 7971.00 & $>19.54$ & $14.96\pm0.02$ & $12.80\pm0.01$ & $10.45\pm0.02$ & SMARTS\\
2017-09-05 & 8002.00 & $>18.80$ & $14.84\pm0.03$ & $12.87\pm0.01$ & $10.35\pm0.03$ & SMARTS\\
2017-10-05 & 8032.00 & $>19.01$ & $14.89\pm0.02$ & $12.71\pm0.01$ & $10.27\pm0.02$ & SMARTS\\
2018-02-01 & 8151.25 & $18.69\pm0.08$ & $14.42\pm0.02$ & $12.37\pm0.01$ & $10.00\pm0.02$ & SMARTS\\
2018-02-12 & 8162.32 & $18.32\pm0.03$&\dots&\dots&\dots& OGLE \\
2018-03-01 & 8179.25 & $18.17\pm0.06$ & $14.18\pm0.02$ & $12.16\pm0.01$ & $9.71\pm0.03$ & SMARTS\\
2018-03-25 & 8202.86 &\dots&\dots&\dots&$9.74\pm0.01$&  VISTA \\
2018-03-30 & 8207.90 &\dots&\dots&\dots&$9.51\pm0.01$&  VISTA \\
2018-04-01 & 8210.25 & $18.10\pm0.07$ & $14.04\pm0.02$ & $12.05\pm0.01$ & $9.62\pm0.02$ & SMARTS\\
2018-04-05 & 8214.32 & $18.12\pm0.04$&\dots&\dots&\dots& OGLE \\
2018-05-01 & 8240.25 & $18.18\pm0.09$ & $14.28\pm0.02$ & $12.25\pm0.01$ & $9.83\pm0.01$ & SMARTS\\ 
2018-05-10 & 8249.73 &\dots&\dots&\dots&$10.12\pm0.01$&  VISTA \\
2018-05-21 & 8249.73 &\dots&\dots&\dots&$9.53\pm0.01$&  VISTA \\
2018-05-24 & 8263.58 &\dots&\dots&\dots&$9.67\pm0.01$&  VISTA \\
2018-05-27 & 8266.56 &\dots&\dots&\dots&$9.81\pm0.01$&  VISTA \\
2018-06-01 & 8271.00 & $18.18\pm0.07$ & $14.17\pm0.02$ & $12.05\pm0.02$ & $9.56\pm0.02$ & SMARTS\\
2018-07-09 & 8309.25 & $17.94\pm0.04$ & $14.05\pm0.02$ & $11.93\pm0.01$ & $9.36\pm0.02$ & SMARTS\\
2018-08-08 & 8338.00 & $18.41\pm0.06$ & $14.47\pm0.02$ & $12.35\pm0.01$ & $9.57\pm0.01$ & SMARTS\\
2018-10-15 & 8406.00 & $18.97\pm0.08$ & $14.90\pm0.02$ & $12.63\pm0.01$ & $9.72\pm0.02$ & SMARTS\\
2019-01-30 & 8514.37 & $18.12\pm0.03$&\dots&\dots&\dots& OGLE \\
2019-02-07 & 8521.25 & $18.23\pm0.04$ & $14.23\pm0.01$ & $11.79\pm0.01$ & $8.86\pm0.02$ & SMARTS\\
2019-02-17 & 8532.33 & $18.42\pm0.05$&\dots&\dots&\dots& OGLE \\
2019-03-06 & 8548.29 & $18.14\pm0.03$&\dots&\dots&\dots& OGLE \\
2019-03-06 & 8548.82 & $18.02\pm0.04$ & $14.06\pm0.01$ & $11.51\pm0.01$ & $8.58\pm0.02$ & SMARTS\\
2019-03-24 & 8566.24 & $18.30\pm0.04$&\dots&\dots&\dots& OGLE \\
2019-04-03 & 8576.20 & $18.34\pm0.04$&\dots&\dots&\dots& OGLE \\
2019-04-06 & 8579.82 & $18.04\pm0.04$ & $14.14\pm0.02$ & $11.63\pm0.01$ & $8.65\pm0.02$ & SMARTS\\
2019-04-11 & 8585.88 & \dots          & \dots          & \dots          & $8.85\pm0.10$ & VISTA\\
2019-04-14 & 8587.15 & $18.18\pm0.04$&\dots&\dots&\dots& OGLE \\
2019-04-22 & 8595.37 & $18.18\pm0.07$&\dots&\dots&\dots& OGLE \\
2019-05-02 & 8606.83 & \dots          & \dots          & \dots          & $8.51\pm0.10$ & VISTA\\
2019-05-07 & 8610.67 & $18.21\pm0.04$ & $14.03\pm0.05$ & $11.41\pm0.02$ & $8.63\pm0.03$ & SMARTS\\
2019-05-14 & 8617.11 & $18.52\pm0.04$&\dots&\dots&\dots& OGLE \\
2019-05-15 & 8618.10 & $18.47\pm0.04$&\dots&\dots&\dots& OGLE \\
2019-06-06 & 8640.04 & $18.69\pm0.05$&\dots&\dots&\dots& OGLE \\
2019-06-08 & 8642.74 & $18.52\pm0.04$ & $14.50\pm0.01$ & $11.63\pm0.01$ & $8.64\pm0.02$ & SMARTS\\
2019-06-15 & 8649.30 & $18.74\pm0.14$&\dots&\dots&\dots& OGLE \\
2019-06-17 & 8652.81 & \dots   & \dots    &  \dots       & $8.58\pm0.10$ & VISTA\\
2019-06-27 & 8661.00 & $18.53\pm0.06$&\dots&\dots&\dots& OGLE \\
2019-07-09 & 8673.65 & $18.57\pm0.04$ & $14.52\pm0.02$ & $11.45\pm0.01$ & $8.47\pm0.01$ & SMARTS\\
2019-08-08 & 8703.14 & \dots & \dots & \dots & $8.761\pm0.10$ & VISTA\\
2019-08-12 & 8707.01 & $19.03\pm0.14$&\dots&\dots&\dots& OGLE \\
2019-08-22 & 8717.08 & \dots & \dots & \dots & $8.673\pm0.10$ & VISTA\\
2019-08-24 & 8719.07 & \dots & \dots & \dots & $8.754\pm0.10$ & VISTA\\
2019-08-25 & 8720.00 & \dots & \dots & \dots & $8.709\pm0.10$ & VISTA\\
2019-08-26 & 8721.08 & \dots & \dots & \dots & $8.597\pm0.10$ & VISTA\\
2019-08-31 & 8726.06 & \dots & \dots & \dots & $8.720\pm0.10$ & VISTA\\
2019-09-08 & 8734.06 & $19.02\pm0.10$&\dots&\dots&\dots& OGLE \\
\enddata
\tablenotetext{a}{From \citet{kospal2017a}}
\end{deluxetable*}

\section{Spectral lines in V346~Nor}

Tabs.~\ref{tab:h2} and \ref{tab:lines} contain the molecular H$_2$ and atomic lines we identified in the X-SHOOTER spectrum of V346~Nor obtained in 2015 June-July. The laboratory and observed wavelengths are given in air.

\startlongtable
\begin{deluxetable*}{ccccccc}
\tablecaption{H$_2$ lines in the spectrum of V346~Nor.\label{tab:h2}}
\tablehead{
\colhead{Line name} & \colhead{Lab. $\lambda$} & \colhead{Obs. $\lambda$} & 
\colhead{Line flux} & \colhead{$A$} & \colhead{$g_J$} & \colhead{$E_J$}\\
\colhead{} & \colhead{(nm)} & \colhead{(nm)} & \colhead{(erg\,s$^{-1}$\,cm$^{-2}$)} & \colhead{(s$^{-1}$)} & \colhead{} & \colhead{(K)} }
\startdata
2$-$0 S(9)     & 1053.27	   & 1053.24 & 6.27e$-$17 $\pm$ 2.06e$-$18 & 3.12e$-$07 & 69 & 20853 \\  
2$-$0 S(8)     & 1057.31	   & 1057.31 & 3.89e$-$17 $\pm$ 2.31e$-$18 & 3.30e$-$07 & 21 & 19434 \\
2$-$0 S(7)     & 1063.84	   & 1063.82 & 1.42e$-$16 $\pm$ 3.66e$-$18 & 3.40e$-$07 & 57 & 18107 \\  
2$-$0 S(6)     & 1072.98	   & 1072.93 & 4.46e$-$17 $\pm$ 2.09e$-$18 & 3.39e$-$07 & 17 & 16880 \\  
2$-$0 S(5)     & 1084.83	   & 1084.80 & 2.66e$-$16 $\pm$ 3.15e$-$18 & 3.28e$-$07 & 45 & 15763 \\  
2$-$0 S(4)     & 1099.52	   & 1099.48 & 1.08e$-$16 $\pm$ 3.17e$-$18 & 3.07e$-$07 & 13 & 14764 \\  
2$-$0 S(3)     & 1117.17	   & 1117.13 & 3.59e$-$16 $\pm$ 1.11e$-$17 & 2.77e$-$07 & 33 & 13890 \\  
2$-$0 S(2)     & 1137.93	   & 1137.88 & 1.15e$-$16 $\pm$ 8.64e$-$18 & 2.38e$-$07 & 9  & 13150 \\  
2$-$0 S(1)     & 1161.90	   & 1161.85 & 2.91e$-$16 $\pm$ 6.26e$-$18 & 1.90e$-$07 & 21 & 12550 \\  
2$-$0 S(0)     & 1189.24	   & 1189.19 & 5.67e$-$17 $\pm$ 3.23e$-$18 & 1.27e$-$07 & 5  & 12095 \\  
2$-$0 Q(1)     & 1238.00	   & 1237.96 & 1.79e$-$16 $\pm$ 3.30e$-$18 & 1.94e$-$07 & 9  & 11789 \\  
2$-$0 Q(2)     & 1241.59	   & 1241.54 & 8.70e$-$17 $\pm$ 2.91e$-$18 & 1.38e$-$07 & 5  & 12095 \\  
2$-$0 Q(3)     & 1246.98	   & 1246.93 & 2.05e$-$16 $\pm$ 3.27e$-$18 & 1.29e$-$07 & 21 & 12550 \\  
2$-$0 Q(4)     & 1254.19	   & 1254.14 & 7.82e$-$17 $\pm$ 2.85e$-$18 & 1.25e$-$07 & 9  & 13150 \\  
2$-$0 Q(5)     & 1263.23	   & 1263.16 & 1.38e$-$16 $\pm$ 6.55e$-$18 & 1.23e$-$07 & 33 & 13890 \\  
2$-$0 Q(6)     & 1274.14	   & 1274.14 & 4.16e$-$17 $\pm$ 6.13e$-$18 & 1.21e$-$07 & 13 & 14764 \\  
2$-$0 Q(7)     & 1286.95	   & 1286.89 & 1.75e$-$16 $\pm$ 4.54e$-$18 & 1.20e$-$07 & 45 & 15763 \\  
2$-$0 O(3)     & 1335.05	   & 1335.00 & 1.91e$-$16 $\pm$ 5.67e$-$18 & 1.61e$-$07 & 9  & 11789 \\  
2$-$0 O(2)     & 1292.87	   & 1292.83 & 5.87e$-$17 $\pm$ 8.60e$-$18 & 3.47e$-$07 & 1  & 11635 \\  
2$-$1 S(3)     & 2072.90	   & 2072.71 & 1.32e$-$15 $\pm$ 2.83e$-$17 & 5.77e$-$07 & 33 & 13890 \\  
2$-$1 S(2)     & 2153.62	   & 2153.43 & 1.02e$-$15 $\pm$ 2.38e$-$17 & 5.60e$-$07 & 9  & 13150 \\  
2$-$1 S(1)     & 2247.11	   & 2246.91 & 1.04e$-$15 $\pm$ 6.67e$-$17 & 4.98e$-$07 & 21 & 12550 \\  
1$-$0 S(9)     & 1687.31	   & 1687.18 & 5.54e$-$16 $\pm$ 7.25e$-$18 & 1.68e$-$07 & 69 & 15722 \\  
1$-$0 S(8)     & 1714.27	   & 1714.14 & 3.50e$-$16 $\pm$ 1.37e$-$17 & 2.34e$-$07 & 21 & 14221 \\  
1$-$0 S(7)     & 1747.47	   & 1747.38 & 1.89e$-$15 $\pm$ 1.66e$-$17 & 2.98e$-$07 & 57 & 12817 \\  
1$-$0 S(6)     & 1787.55	   & 1787.53 & 1.50e$-$15 $\pm$ 1.27e$-$16 & 3.54e$-$07 & 17 & 11522 \\  
1$-$0 S(3)     & 1957.03	   & 1956.87 & 1.62e$-$14 $\pm$ 3.34e$-$16 & 4.21e$-$07 & 33 & 8365  \\  
1$-$0 S(2)     & 2033.20	   & 2033.02 & 4.48e$-$15 $\pm$ 1.14e$-$16 & 3.98e$-$07 & 9  & 7584  \\  
1$-$0 S(1)     & 2121.25	   & 2121.08 & 1.33e$-$14 $\pm$ 2.59e$-$16 & 3.47e$-$07 & 21 & 6956  \\  
1$-$0 S(0)     & 2222.68	   & 2222.52 & 3.01e$-$15 $\pm$ 6.58e$-$17 & 2.53e$-$07 & 5  & 6471  \\  
1$-$0 Q(1)     & 2405.93	   & 2405.76 & 1.21e$-$14 $\pm$ 2.81e$-$16 & 4.29e$-$07 & 9  & 6149  \\  
1$-$0 Q(2)     & 2412.78	   & 2412.62 & 2.86e$-$15 $\pm$ 1.11e$-$16 & 3.03e$-$07 & 5  & 6471  \\  
1$-$0 Q(3)     & 2423.07	   & 2422.89 & 8.56e$-$15 $\pm$ 1.71e$-$16 & 2.78e$-$07 & 21 & 6956  \\  
1$-$0 Q(4)     & 2436.82	   & 2436.69 & 2.09e$-$15 $\pm$ 8.22e$-$17 & 2.65e$-$07 & 9  & 7586  \\  
1$-$0 Q(5)     & 2454.08	   & 2454.04 & 9.88e$-$16 $\pm$ 3.24e$-$16 & 2.55e$-$07 & 33 & 8365  \\  
1$-$0 Q(6)     & 2474.87	   & 2475.04 & 6.17e$-$16 $\pm$ 1.02e$-$16 & 2.45e$-$07 & 13 & 9286  \\  
3$-$1 S(7)     & 1130.11	   & 1130.07 & 3.89e$-$17 $\pm$ 7.38e$-$18 & 7.56e$-$07 & 57 & 23069 \\  
3$-$1 S(6)     & 1139.35	   & 1139.32 & 2.22e$-$17 $\pm$ 4.07e$-$18 & 7.71e$-$07 & 17 & 21911 \\  
3$-$1 S(5)     & 1151.54	   & 1151.50 & 1.83e$-$16 $\pm$ 9.09e$-$18 & 7.62e$-$07 & 45 & 20856 \\  
3$-$1 S(4)     & 1166.84	   & 1166.79 & 6.31e$-$17 $\pm$ 2.72e$-$18 & 7.27e$-$07 & 13 & 19912 \\  
3$-$1 S(3)     & 1185.37	   & 1185.32 & 1.63e$-$16 $\pm$ 4.34e$-$18 & 6.67e$-$07 & 33 & 19086 \\  
3$-$1 S(2)     & 1207.26	   & 1207.19 & 3.59e$-$17 $\pm$ 2.26e$-$18 & 5.83e$-$07 & 9  & 18386 \\  
3$-$1 S(1)     & 1232.66	   & 1232.60 & 1.29e$-$16 $\pm$ 1.69e$-$18 & 4.73e$-$07 & 21 & 17818 \\  
3$-$1 S(0)     & 1261.72	   & 1261.66 & 1.82e$-$17 $\pm$ 3.28e$-$18 & 3.12e$-$07 & 5  & 17387 \\
\enddata
\end{deluxetable*}

\startlongtable
\begin{deluxetable*}{cccccccc}
\tablecaption{Atomic lines in the spectrum of V346~Nor.\label{tab:lines}}
\tablehead{
\colhead{Species} & \colhead{Lab. $\lambda$} & \colhead{Obs. $\lambda$} & 
\colhead{Line flux} & \colhead{Lower level} & \colhead{Upper level} & \colhead{$A_{ki}$} & \colhead{$E_i - E_k$}\\
\colhead{} & \colhead{(nm)} & \colhead{(nm)} & \colhead{(erg\,s$^{-1}$\,cm$^{-2}$)} & \colhead{Conf., term, J} & \colhead{Conf., term, J} & \colhead{(s$^{-1}$)} & \colhead{(eV)}}
\startdata
HI           & 656.279     & 656.180 & 2.43e$-$16$\pm$1.87e$-$18                    & 2                                         & 3                                         & 4.41e+7   & 10.199 --	12.088 \\ \relax
[CI]	     & 872.713     & 872.660 & 3.81e$-$17$\pm$1.06e$-$18                    & 2s$^2$2p$^2$ $^1$D 2                      & 2s$^2$2p$^2$ $^1$S 0                      & 6.0e$-$1  & 1.264 -- 2.684 \\ \relax
[CI]	     & 982.412     & 982.240 & 1.43e$-$16$\pm$2.85e$-$18                    & 2s$^2$2p$^2$ $^3$P 1                      & 2s$^2$2p$^2$ $^1$D 2                      & 7.3e$-$5  & 0.002 -- 1.264 \\ \relax
[CI]	     & 985.025     & 983.680 & 4.95e$-$16$\pm$2.43e$-$18                    & 2s$^2$2p$^2$ $^3$P 2                      & 2s$^2$2p$^2$ $^1$D 2                      & 2.2e$-$4  & 0.005 -- 1.264 \\ \relax
[NI]	     & 1039.77     & 1039.55 & 3.00e$-$16$\pm$1.64e$-$17                     & 2s$^2$2p$^3$ $^2$D$^{\circ}$ \sfrac{5}{2} & 2s$^2$2p$^3$ $^2$P$^{\circ}$ \sfrac{3}{2} & 6.0e$-$2  & 2.384 -- 3.576 \\ \relax
[NI]	     & 1039.82     & 1039.75 & 3.00e$-$16$\pm$1.64e$-$17                     & 2s$^2$2p$^3$ $^2$D$^{\circ}$ \sfrac{5}{2} & 2s$^2$2p$^3$ $^2$P$^{\circ}$ \sfrac{1}{2} & 3.45e$-$2 & 2.384 -- 3.576 \\ \relax
[NI]	     & 1040.72     & 1040.51 & 1.90e$-$16$\pm$1.64e$-$17                     & 2s$^2$2p$^3$ $^2$D$^{\circ}$ \sfrac{3}{2} & 2s$^2$2p$^3$ $^2$P$^{\circ}$ \sfrac{3}{2} & 2.56e$-$2 & 2.385 -- 3.576 \\ \relax
[NI]	     & 1040.76     & 1040.71 & 1.90e$-$16$\pm$1.64e$-$17                     & 2s$^2$2p$^3$ $^2$D$^{\circ}$ \sfrac{3}{2} & 2s$^2$2p$^3$ $^2$P$^{\circ}$ \sfrac{1}{2} & 5.2e$-$2  & 2.385 -- 3.576 \\ \relax
[NII]        & 654.805     & 654.680 & 1.90e$-$17$\pm$1.93e$-$18                    & 2s$^2$2p$^2$ $^3$P 1                      & 2s$^2$2p$^2$ $^1$D 2                      & 9.84e$-$4 & 0.006 -- 1.899 \\ \relax
[NII]	     & 658.345     & 656.220 & 5.74e$-$17$\pm$1.70e$-$18                    & 2s$^2$2p$^2$ $^3$P 2                      & 2s$^2$2p$^2$ $^1$D 2                      & 2.91e$-$3 & 0.016 -- 1.899 \\ \relax
[OI]	     & 630.030     & 629.900 & 4.66e$-$16$\pm$1.20e$-$18                    & 2s$^2$2p$^4$ $^3$P 2                      & 2s$^2$2p$^4$ $^1$D 2                      & 5.63e$-$3 & 0.000 -- 1.967 \\ \relax
[OI]	     & 636.378     & 636.260 & 1.79e$-$16$\pm$1.18e$-$18                    & 2s$^2$2p$^4$ $^3$P 1                      & 2s$^2$2p$^4$ $^1$D 2                      & 1.82e$-$3 & 0.020 -- 1.967 \\ \relax
[SII]	     & 671.644     & 671.520 & 3.00e$-$16$\pm$1.73e$-$18                    & 3s$^2$3p$^3$ $^4$S$^{\circ}$ \sfrac{3}{2} & 3s$^2$3p$^3$ $^2$D$^{\circ}$ \sfrac{5}{2} & 1.88e$-$4 & 0.000 -- 1.845 \\ \relax
[SII]	     & 673.082     & 672.960 & 5.51e$-$16$\pm$1.68e$-$18                    & 3s$^2$3p$^3$ $^4$S$^{\circ}$ \sfrac{3}{2} & 3s$^2$3p$^3$ $^2$D$^{\circ}$ \sfrac{3}{2} & 1.21e$-$4 & 0.000 -- 1.842 \\ \relax
[SII]	     & 1028.67	   & 1028.52 & 3.00e$-$16$\pm$2.00e$-$17                    & 3s$^2$3p$^3$ $^2$D$^{\circ}$ \sfrac{3}{2} & 3s$^2$3p$^3$ $^2$P$^{\circ}$ \sfrac{3}{2} & 5.25e$-$2 & 1.842 -- 3.046 \\ \relax
[SII]	     & 1032.05	   & 1031.82 & 4.25e$-$16$\pm$1.03e$-$17                    & 3s$^2$3p$^3$ $^2$D$^{\circ}$ \sfrac{5}{2} & 3s$^2$3p$^3$ $^2$P$^{\circ}$ \sfrac{3}{2} & 1.22e$-$1 & 1.845 -- 3.046 \\ \relax
[SII]	     & 1033.64	   & 1033.44 & 3.04e$-$16$\pm$1.57e$-$17                    & 3s$^2$3p$^3$ $^2$D$^{\circ}$ \sfrac{3}{2} & 3s$^2$3p$^3$ $^2$P$^{\circ}$ \sfrac{1}{2} & 1.04e$-$1 & 1.842 -- 3.041 \\ \relax
[SII]	     & 1037.05	   & 1036.80 & 1.25e$-$16$\pm$6.08e$-$18                    & 3s$^2$3p$^3$ $^2$D$^{\circ}$ \sfrac{5}{2} & 3s$^2$3p$^3$ $^2$P$^{\circ}$ \sfrac{1}{2} & 6.81e$-$2 & 1.845 -- 3.041 \\ \relax
[CaII]	     & 729.147     & 729.020 & 1.68e$-$16$\pm$1.31e$-$18                    & 3p$^6$4s $^2$S \sfrac{1}{2}               & 3p$^6$3d $^2$D \sfrac{5}{2}               & 1.3e+0    & 0.000 -- 1.700 \\ \relax
[CaII]	     & 732.389     & 732.260 & 1.16e$-$16$\pm$1.05e$-$18                    & 3p$^6$4s $^2$S \sfrac{1}{2}               & 3p$^6$3d $^2$D \sfrac{3}{2}               & 1.3e+0    & 0.000 -- 1.692 \\ \relax
CaII         & 849.802     & 849.620 & 8.87e$-$18$\pm$2.33e$-$19\tablenotemark{$a$} & 3p$^6$3d $^2$D \sfrac{3}{2}               & 3p$^6$4p $^2$P$^{\circ}$ \sfrac{3}{2}     & 1.11e+6   & 1.692 -- 3.151 \\ \relax
             &             & 849.980 & 2.80e$-$17$\pm$2.24e$-$18\tablenotemark{$b$} &                                           &                                           &                            \\ \relax
CaII         & 854.209     & 854.040 & 4.90e$-$17$\pm$2.01e$-$19\tablenotemark{$a$} & 3p$^6$3d $^2$D \sfrac{5}{2}               & 3p$^6$4p $^2$P$^{\circ}$ \sfrac{3}{2}     & 9.9e+6    & 1.700 -- 3.151 \\ \relax
             &             & 854.445 & 3.87e$-$17$\pm$3.71e$-$18\tablenotemark{$b$} &                                           &                                           &                            \\ \relax
CaII         & 866.214     & 866.040 & 2.94e$-$17$\pm$3.07e$-$19\tablenotemark{$a$} & 3p$^6$3d $^2$D \sfrac{3}{2}               & 3p$^6$4p $^2$P$^{\circ}$ \sfrac{1}{2}     & 1.06e+7   & 1.692 -- 3.123 \\ \relax 
             &             & 866.445 & 3.09e$-$17$\pm$2.55e$-$18\tablenotemark{$b$} &                                           &                                           &                            \\ \relax
[VI]	     & 1188.31     & 1188.06 & 2.44e$-$16$\pm$1.41e$-$17                    & 3d$^3$4s$^2$ a $^4$F \sfrac{3}{2}         & 3d$^4$($^5$D)4s a $^4$D \sfrac{1}{2}      & 4.2e$-$2  & 0.000 -- 1.043 \\ \relax
[CrII]	     & 800.007     & 799.840 & 2.41e$-$17$\pm$8.60e$-$19                    & 3d$^5$ a $^6$S \sfrac{5}{2}               & 3d$^4$($^5$D)4s a $^6$D \sfrac{9}{2}      & 1.0e$-$1  & 0.000 -- 1.549 \\ \relax
[CrII]	     & 812.532     & 812.380 & 2.09e$-$17$\pm$7.26e$-$19                    & 3d$^5$ a $^6$S \sfrac{5}{2}               & 3d$^4$($^5$D)4s a $^6$D \sfrac{7}{2}      & 9.4e$-$2  & 0.000 -- 1.525 \\ \relax
[CrII]	     & 822.970     & 822.800 & 1.31e$-$17$\pm$7.45e$-$19                    & 3d$^5$ a $^6$S \sfrac{5}{2}               & 3d$^4$($^5$D)4s a $^6$D \sfrac{5}{2}      & 8.8e$-$2  & 0.000 -- 1.506 \\ \relax
[CrII]	     & 830.851     & 830.680 & 7.07e$-$18$\pm$1.44e$-$18                    & 3d$^5$ a $^6$S \sfrac{5}{2}               & 3d$^4$($^5$D)4s a $^6$D \sfrac{3}{2}      & 8.4e$-$2  & 0.000 -- 1.492 \\ \relax
[CrII]	     & 835.769     & 835.580 & 5.63e$-$18$\pm$1.02e$-$18                    & 3d$^5$ a $^6$S \sfrac{5}{2}               & 3d$^4$($^5$D)4s a $^6$D \sfrac{1}{2}      & 8.2e$-$2  & 0.000 -- 1.483 \\ \relax
[FeII]	     & 715.517     & 715.380 & 1.52e$-$16$\pm$8.52e$-$19                    & 3d$^7$ a $^4$F \sfrac{9}{2}               & 3d$^7$ a $^2$G \sfrac{9}{2}               & 1.46e$-$1 & 0.232 -- 1.964 \\ \relax
[FeII]	     & 717.200     & 717.060 & 3.98e$-$17$\pm$8.96e$-$19                    & 3d$^7$ a $^4$F \sfrac{7}{2}               & 3d$^7$ a $^2$G \sfrac{7}{2}               & 5.51e$-$2 & 0.301 -- 2.030 \\ \relax
[FeII]	     & 738.817     & 738.660 & 3.69e$-$17$\pm$1.27e$-$18                    & 3d$^7$ a $^4$F \sfrac{5}{2}               & 3d$^7$ a $^2$G \sfrac{7}{2}               & 4.21e$-$2 & 0.352 -- 2.030 \\ \relax
[FeII]	     & 745.256     & 745.100 & 6.32e$-$17$\pm$8.50e$-$19                    & 3d$^7$ a $^4$F \sfrac{7}{2}               & 3d$^7$ a $^2$G \sfrac{9}{2}               & 4.77e$-$2 & 0.301 -- 1.964 \\ \relax
[FeII]	     & 763.752     & 763.600 & 1.30e$-$17$\pm$1.04e$-$18                    & 3d$^6$($^5$D)4s a $^6$D \sfrac{7}{2}      & 3d$^7$ a $^4$P \sfrac{5}{2}               & 6.6e$-$3  & 0.048 -- 1.671 \\ \relax
[FeII]       & 766.528     & 766.300 & 4.69e$-$18$\pm$1.00e$-$18                    & 3d$^6$($^5$D)4s a $^6$D \sfrac{3}{2}      & 3d$^7$ a $^4$P \sfrac{1}{2}               & 6.2e$-$3  & 0.107 -- 1.724 \\ \relax
[FeII]	     & 768.693     & 768.580 & 2.03e$-$17$\pm$9.59e$-$19                    & 3d$^6$($^5$D)4s a $^6$D \sfrac{5}{2}      & 3d$^7$ a $^4$P \sfrac{3}{2}               & 6.8e$-$3  & 0.083 -- 1.695 \\ \relax
[FeII]       & 773.313     & 773.060 & 1.19e$-$18$\pm$9.00e$-$19                    & 3d$^6$($^5$D)4s a $^6$D \sfrac{1}{2}      & 3d$^7$ a $^4$P \sfrac{1}{2}               & 1.93e$-$3 & 0.121 -- 1.724 \\ \relax
[FeII]	     & 861.695     & 861.520 & 3.58e$-$16$\pm$1.21e$-$18                    & 3d$^7$ a $^4$F \sfrac{9}{2}               & 3d$^7$ a $^4$P \sfrac{5}{2}               & 3.6e$-$2  & 0.232 -- 1.671 \\ \relax
[FeII]	     & 889.193     & 889.040 & 1.68e$-$16$\pm$1.21e$-$18                    & 3d$^7$ a $^4$F \sfrac{7}{2}               & 3d$^7$ a $^4$P \sfrac{3}{2}               & 2.21e$-$2 & 0.301 -- 1.695 \\ \relax
[FeII]	     & 903.349     & 903.100 & 5.41e$-$17$\pm$1.40e$-$18                    & 3d$^7$ a $^4$F \sfrac{5}{2}               & 3d$^7$ a $^4$P \sfrac{1}{2}               & 1.61e$-$2 & 0.352 -- 1.724 \\ \relax
[FeII]	     & 905.195     & 905.020 & 1.11e$-$16$\pm$1.19e$-$18                    & 3d$^7$ a $^4$F \sfrac{7}{2}               & 3d$^7$ a $^4$P \sfrac{5}{2}               & 8.8e$-$3  & 0.301 -- 1.671 \\ \relax
[FeII]	     & 922.663     & 922.500 & 1.13e$-$16$\pm$1.66e$-$18                    & 3d$^7$ a $^4$F \sfrac{5}{2}               & 3d$^7$ a $^4$P \sfrac{3}{2}               & 1.3e$-$2  & 0.352 -- 1.695 \\ \relax
[FeII]	     & 926.755     & 926.560 & 8.22e$-$17$\pm$1.46e$-$18                    & 3d$^7$ a $^4$F \sfrac{3}{2}               & 3d$^7$ a $^4$P \sfrac{1}{2}               & 2.1e$-$2  & 0.387 -- 1.724 \\ \relax
[FeII]	     & 939.904     & 939.740 & 3.20e$-$17$\pm$2.02e$-$18                    & 3d$^7$ a $^4$F \sfrac{5}{2}               & 3d$^7$ a $^4$P \sfrac{5}{2}               & 1.7e$-$3  & 0.352 -- 1.671 \\ \relax
[FeII]       & 947.094     & 947.020 & 3.38e$-$17$\pm$4.88e$-$18                    & 3d$^7$ a $^4$F \sfrac{3}{2}               & 3d$^7$ a $^4$P \sfrac{3}{2}               & 3.7e$-$3  & 0.387 -- 1.695 \\ \relax
[FeII]	     & 1256.68     & 1256.40 & 3.39e$-$15$\pm$1.47e$-$17                    & 3d$^6$($^5$D)4s a $^6$D \sfrac{9}{2}      & 3d$^6$($^5$D)4s a $^4$D \sfrac{7}{2}      & 4.74e$-$3 & 0.000 -- 0.986 \\ \relax
[FeII]	     & 1270.34     & 1270.08 & 4.52e$-$16$\pm$4.88e$-$17                    & 3d$^6$($^5$D)4s a $^6$D \sfrac{1}{2}      & 3d$^6$($^5$D)4s a $^4$D \sfrac{1}{2}      & 3.32e$-$3 & 0.121 -- 1.097 \\ \relax
[FeII]	     & 1278.77     & 1278.48 & 5.63e$-$16$\pm$3.06e$-$17                    & 3d$^6$($^5$D)4s a $^6$D \sfrac{3}{2}      & 3d$^6$($^5$D)4s a $^4$D \sfrac{3}{2}      & 2.45e$-$3 & 0.107 -- 1.076 \\ \relax
[FeII]	     & 1294.27     & 1294.02 & 7.35e$-$16$\pm$1.49e$-$17                    & 3d$^6$($^5$D)4s a $^6$D \sfrac{5}{2}      & 3d$^6$($^5$D)4s a $^4$D \sfrac{5}{2}      & 1.98e$-$3 & 0.083 -- 1.040 \\ \relax
[FeII]	     & 1297.77     & 1297.50 & 3.06e$-$16$\pm$1.10e$-$17                    & 3d$^6$($^5$D)4s a $^6$D \sfrac{1}{2}      & 3d$^6$($^5$D)4s a $^4$D \sfrac{3}{2}      & 1.08e$-$3 & 0.121 -- 1.076 \\ \relax
[FeII]	     & 1320.55     & 1320.30 & 1.13e$-$15$\pm$1.06e$-$17                    & 3d$^6$($^5$D)4s a $^6$D \sfrac{7}{2}      & 3d$^6$($^5$D)4s a $^4$D \sfrac{7}{2}      & 1.31e$-$3 & 0.048 -- 0.986 \\ \relax
[FeII]       & 1327.77     & 1327.50 & 4.64e$-$16$\pm$2.06e$-$17                    & 3d$^6$($^5$D)4s a $^6$D \sfrac{3}{2}      & 3d$^6$($^5$D)4s a $^4$D \sfrac{5}{2}      & 1.17e$-$3 & 0.107 -- 1.040 \\ \relax
[FeII]	     & 1533.47     & 1553.12 & 1.57e$-$15$\pm$5.34e$-$17                    & 3d$^7$ a $^4$F \sfrac{9}{2}               & 3d$^6$($^5$D)4s a $^4$D \sfrac{5}{2}      & 3.12e$-$3 & 0.232 -- 1.040 \\ \relax
[FeII]	     & 1599.48     & 1599.12 & 1.18e$-$15$\pm$2.30e$-$17                    & 3d$^7$ a $^4$F \sfrac{7}{2}               & 3d$^6$($^5$D)4s a $^4$D \sfrac{3}{2}      & 4.18e$-$3 & 0.301 -- 1.076 \\ \relax
[FeII]	     & 1643.55     & 1643.16 & 5.95e$-$15$\pm$4.66e$-$17                    & 3d$^7$ a $^4$F \sfrac{9}{2}               & 3d$^6$($^5$D)4s a $^4$D \sfrac{7}{2}      & 6.0e$-$3  & 0.232 -- 0.986 \\ \relax
[FeII]	     & 1663.77     & 1663.44 & 7.66e$-$16$\pm$2.59e$-$17                    & 3d$^7$ a $^4$F \sfrac{5}{2}               & 3d$^6$($^5$D)4s a $^4$D \sfrac{1}{2}      & 4.75e$-$3 & 0.352 -- 1.097 \\ \relax
[FeII]	     & 1676.88     & 1676.52 & 1.20e$-$15$\pm$3.42e$-$17                    & 3d$^7$ a $^4$F \sfrac{7}{2}               & 3d$^6$($^5$D)4s a $^4$D \sfrac{5}{2}      & 2.49e$-$3 & 0.301 -- 1.040 \\ \relax
[FeII]	     & 1711.13     & 1710.90 & 1.68e$-$16$\pm$2.49e$-$17                    & 3d$^7$ a $^4$F \sfrac{5}{2}               & 3d$^6$($^5$D)4s a $^4$D \sfrac{3}{2}      & 1.18e$-$3 & 0.352 -- 1.076 \\ \relax
[NiII]	     & 737.783	   & 737.660 & 8.45e$-$17$\pm$9.82e$-$19                    & 3p$^6$3d$^9$ $^2$D \sfrac{5}{2}           & 3p$^6$3d$^8$($^3$F)4s $^2$F \sfrac{7}{2}  & 2.3e$-$1  & 0.000 -- 1.680 \\ \relax
[NiII]	     & 741.161	   & 741.080 & 1.68e$-$17$\pm$8.22e$-$19                    & 3p$^6$3d$^9$ $^2$D \sfrac{3}{2}           & 3p$^6$3d$^8$($^3$F)4s $^2$F \sfrac{5}{2}  & 1.8e$-$1  & 0.187 -- 1.859
\enddata
\tablenotetext{a}{Shock component for the Ca IRT.}
\tablenotetext{b}{Gaussian component for the Ca IRT.}
\end{deluxetable*}

\bibliography{paper}{}

\begin{thebibliography}{}
\expandafter\ifx\csname natexlab\endcsname\relax\def\natexlab#1{#1}\fi
\providecommand{\url}[1]{\href{#1}{#1}}

\bibitem[{{Acosta-Pulido} {et~al.}(2007){Acosta-Pulido}, {Kun},
  {{\'A}brah{\'a}m}, {K{\'o}sp{\'a}l}, {Csizmadia}, {Kiss}, {Mo{\'o}r},
  {Szabados}, {Benk{\H o}}, {Barrena Delgado}, {Charcos-Llorens}, {Eredics},
  {Kiss}, {Manchado}, {R{\'a}cz}, {Ramos Almeida}, {Sz{\'e}kely}, \&
  {Vidal-N{\'u}{\~n}ez}}]{acosta2007}
{Acosta-Pulido}, J.~A., {Kun}, M., {{\'A}brah{\'a}m}, P., {et~al.} 2007, \aj,
  133, 2020

\bibitem[{{Aspin} {et~al.}(2006){Aspin}, {Barbieri}, {Boschi}, {Di Mille},
  {Rampazzi}, {Reipurth}, \& {Tsvetkov}}]{aspin2006}
{Aspin}, C., {Barbieri}, C., {Boschi}, F., {et~al.} 2006, \aj, 132, 1298

\bibitem[{{Audard} {et~al.}(2014){Audard}, {{\'A}brah{\'a}m}, {Dunham},
  {Green}, {Grosso}, {Hamaguchi}, {Kastner}, {K{\'o}sp{\'a}l}, {Lodato},
  {Romanova}, {Skinner}, {Vorobyov}, \& {Zhu}}]{audard2014}
{Audard}, M., {{\'A}brah{\'a}m}, P., {Dunham}, M.~M., {et~al.} 2014, Protostars
  and Planets VI, 387

\bibitem[{{Beck} {et~al.}(2008){Beck}, {McGregor}, {Takami}, \&
  {Pyo}}]{beck2008}
{Beck}, T.~L., {McGregor}, P.~J., {Takami}, M., \& {Pyo}, T.-S. 2008, \apj,
  676, 472

\bibitem[{{Black} \& {van Dishoeck}(1987)}]{black1987}
{Black}, J.~H., \& {van Dishoeck}, E.~F. 1987, The Astrophysical Journal, 322,
  412

\bibitem[{{Connelley} \& {Reipurth}(2018)}]{connelley2018}
{Connelley}, M.~S., \& {Reipurth}, B. 2018, The Astrophysical Journal, 861, 145

\bibitem[{{Cotten} \& {Song}(2016)}]{cotten2016}
{Cotten}, T.~H., \& {Song}, I. 2016, \apjs, 225, 15

\bibitem[{{Cutri} {et~al.}(2015){Cutri}, {Mainzer}, {Conrow}, {Masci}, {Bauer},
  {Dailey}, {Kirkpatrick}, {Fajardo-Acosta}, {Gelino}, {Grillmair}, {Wheelock},
  {Yan}, {Harbut}, {Beck}, {Wittman}, {Wright}, {Masiero}, {Grav}, {Sonnett},
  {Nugent}, {Kramer}, {Stevenson}, {Eisenhardt}, {Fabinsky}, {Tholen}, {Papin},
  {Fowler}, \& {McCallon}}]{cutri2015}
{Cutri}, R.~M., {Mainzer}, A., {Conrow}, T., {et~al.} 2015, {Explanatory
  Supplement to the NEOWISE Data Release Products}, Tech. rep.

\bibitem[{{Davis} {et~al.}(2003){Davis}, {Whelan}, {Ray}, \&
  {Chrysostomou}}]{davis2003}
{Davis}, C.~J., {Whelan}, E., {Ray}, T.~P., \& {Chrysostomou}, A. 2003, \aap,
  397, 693

\bibitem[{{Dere} {et~al.}(2019){Dere}, {Del Zanna}, {Young}, {Landi}, \&
  {Sutherland}}]{dere2019}
{Dere}, K.~P., {Del Zanna}, G., {Young}, P.~R., {Landi}, E., \& {Sutherland},
  R.~S. 2019, The Astrophysical Journal Supplement Series, 241, 22

\bibitem[{{Dere} {et~al.}(1997){Dere}, {Landi}, {Mason}, {Monsignori Fossi}, \&
  {Young}}]{dere1997}
{Dere}, K.~P., {Landi}, E., {Mason}, H.~E., {Monsignori Fossi}, B.~C., \&
  {Young}, P.~R. 1997, Astronomy and Astrophysics Supplement Series, 125, 149

\bibitem[{{Dunham} {et~al.}(2013){Dunham}, {Arce}, {Allen}, {Evans},
  {Broekhoven-Fiene}, {Chapman}, {Cieza}, {Gutermuth}, {Harvey}, {Hatchell},
  {Huard}, {Kirk}, {Matthews}, {Mer{\'{\i}}n}, {Miller}, {Peterson}, \&
  {Spezzi}}]{dunham2013}
{Dunham}, M.~M., {Arce}, H.~G., {Allen}, L.~E., {et~al.} 2013, \aj, 145, 94

\bibitem[{{Dunham} {et~al.}(2014){Dunham}, {Stutz}, {Allen}, {Evans},
  {Fischer}, {Megeath}, {Myers}, {Offner}, {Poteet}, {Tobin}, \&
  {Vorobyov}}]{dunham2014ppvi}
{Dunham}, M.~M., {Stutz}, A.~M., {Allen}, L.~E., {et~al.} 2014, Protostars and
  Planets VI, 195

\bibitem[{{Eisl{\"o}ffel} \& {Mundt}(1997)}]{eisloffel1997}
{Eisl{\"o}ffel}, J., \& {Mundt}, R. 1997, \aj, 114, 280

\bibitem[{{Eisl{\"o}ffel} {et~al.}(2000){Eisl{\"o}ffel}, {Smith}, \&
  {Davis}}]{eisloffel2000}
{Eisl{\"o}ffel}, J., {Smith}, M.~D., \& {Davis}, C.~J. 2000, \aap, 359, 1147

\bibitem[{{Elias}(1980)}]{elias1980}
{Elias}, J.~H. 1980, \apj, 241, 728

\bibitem[{{Giannini} {et~al.}(2018){Giannini}, {Lorenzetti}, {Antoniucci},
  {Boschin}, \& {Harutyunyan}}]{giannini2018}
{Giannini}, T., {Lorenzetti}, D., {Antoniucci}, S., {Boschin}, W., \&
  {Harutyunyan}, A. 2018, The Astronomer's Telegram, 12054, 1

\bibitem[{{Graham} \& {Frogel}(1985)}]{graham1985}
{Graham}, J.~A., \& {Frogel}, J.~A. 1985, \apj, 289, 331

\bibitem[{{Hartmann} \& {Kenyon}(1996)}]{hk96}
{Hartmann}, L., \& {Kenyon}, S.~J. 1996, \araa, 34, 207

\bibitem[{{Hollenbach}(1997)}]{hollenbach1997}
{Hollenbach}, D. 1997, Herbig-Haro Flows and the Birth of Stars; IAU Symposium
  No. 182, 182, 181

\bibitem[{{Kausch} {et~al.}(2015){Kausch}, {Noll}, {Smette}, {Kimeswenger},
  {Barden}, {Szyszka}, {Jones}, {Sana}, {Horst}, \& {Kerber}}]{kausch2015}
{Kausch}, W., {Noll}, S., {Smette}, A., {et~al.} 2015, \aap, 576, A78

\bibitem[{{Kemper} {et~al.}(2004){Kemper}, {Vriend}, \& {Tielens}}]{kemper2004}
{Kemper}, F., {Vriend}, W.~J., \& {Tielens}, A.~G.~G.~M. 2004, \apj, 609, 826

\bibitem[{{K{\'o}sp{\'a}l} {et~al.}(2017{\natexlab{a}}){K{\'o}sp{\'a}l},
  {{\'A}brah{\'a}m}, {Westhues}, \& {Haas}}]{kospal2017a}
{K{\'o}sp{\'a}l}, {\'A}., {{\'A}brah{\'a}m}, P., {Westhues}, C., \& {Haas}, M.
  2017{\natexlab{a}}, \aap, 597, L10

\bibitem[{{K{\'o}sp{\'a}l} {et~al.}(2017{\natexlab{b}}){K{\'o}sp{\'a}l},
  {{\'A}brah{\'a}m}, {Csengeri}, {Feh{\'e}r}, {Hogerheijde}, {Brinch},
  {Dunham}, {Vorobyov}, {Salter}, \& {Henning}}]{kospal2017c}
{K{\'o}sp{\'a}l}, {\'A}., {{\'A}brah{\'a}m}, P., {Csengeri}, T., {et~al.}
  2017{\natexlab{b}}, \apj, 843, 45

\bibitem[{{Kraus} {et~al.}(2016){Kraus}, {Caratti o Garatti}, {Garcia-Lopez},
  {Kreplin}, {Aarnio}, {Monnier}, {Naylor}, \& {Weigelt}}]{kraus2016}
{Kraus}, S., {Caratti o Garatti}, A., {Garcia-Lopez}, R., {et~al.} 2016,
  \mnras, 462, L61

\bibitem[{{Lagage} {et~al.}(2004){Lagage}, {Pel}, {Authier}, {Belorgey},
  {Claret}, {Doucet}, {Dubreuil}, {Durand}, {Elswijk}, {Girardot}, {K{\"a}ufl},
  {Kroes}, {Lortholary}, {Lussignol}, {Marchesi}, {Pantin}, {Peletier},
  {Pirard}, {Pragt}, {Rio}, {Schoenmaker}, {Siebenmorgen}, {Silber}, {Smette},
  {Sterzik}, \& {Veyssiere}}]{lagage2004}
{Lagage}, P.~O., {Pel}, J.~W., {Authier}, M., {et~al.} 2004, The Messenger,
  117, 12

\bibitem[{{Mainzer} {et~al.}(2014){Mainzer}, {Bauer}, {Cutri}, {Grav},
  {Masiero}, {Beck}, {Clarkson}, {Conrow}, {Dailey}, {Eisenhardt}, {Fabinsky},
  {Fajardo-Acosta}, {Fowler}, {Gelino}, {Grillmair}, {Heinrichsen}, {Kendall},
  {Kirkpatrick}, {Liu}, {McCallon}, {Nugent}, {Papin}, {Rice}, {Royer}, {Ryan},
  {Sevilla}, {Sonnett}, {Stevenson}, {Thompson}, {Wheelock}, {Wiemer},
  {Wittman}, {Wright}, \& {Yan}}]{mainzer2014}
{Mainzer}, A., {Bauer}, J., {Cutri}, R.~M., {et~al.} 2014, \apj, 792, 30

\bibitem[{{Mathis}(1990)}]{mathis1990}
{Mathis}, J.~S. 1990, \araa, 28, 37

\bibitem[{{Minniti} {et~al.}(2010){Minniti}, {Lucas}, {Emerson}, {Saito},
  {Hempel}, {Pietrukowicz}, {Ahumada}, {Alonso}, {Alonso-Garcia}, {Arias},
  {Bandyopadhyay}, {Barb{\'a}}, {Barbuy}, {Bedin}, {Bica}, {Borissova},
  {Bronfman}, {Carraro}, {Catelan}, {Clari{\'a}}, {Cross}, {de Grijs},
  {D{\'e}k{\'a}ny}, {Drew}, {Fari{\~n}a}, {Feinstein}, {Fern{\'a}ndez
  Laj{\'u}s}, {Gamen}, {Geisler}, {Gieren}, {Goldman}, {Gonzalez}, {Gunthardt},
  {Gurovich}, {Hambly}, {Irwin}, {Ivanov}, {Jord{\'a}n}, {Kerins}, {Kinemuchi},
  {Kurtev}, {L{\'o}pez-Corredoira}, {Maccarone}, {Masetti}, {Merlo},
  {Messineo}, {Mirabel}, {Monaco}, {Morelli}, {Padilla}, {Palma}, {Parisi},
  {Pignata}, {Rejkuba}, {Roman-Lopes}, {Sale}, {Schreiber}, {Schr{\"o}der},
  {Smith}, {}, {Soto}, {Tamura}, {Tappert}, {Thompson}, {Toledo}, {Zoccali}, \&
  {Pietrzynski}}]{minniti2010}
{Minniti}, D., {Lucas}, P.~W., {Emerson}, J.~P., {et~al.} 2010, \na, 15, 433

\bibitem[{{Molinari} {et~al.}(1993){Molinari}, {Liseau}, {Lorenzetti}, \&
  {Graham}}]{molinari1993}
{Molinari}, S., {Liseau}, R., {Lorenzetti}, D., \& {Graham}, J. 1993, \iaucirc,
  5727

\bibitem[{{Natta} \& {Whitney}(2000)}]{natta2000}
{Natta}, A., \& {Whitney}, B.~A. 2000, \aap, 364, 633

\bibitem[{{Ninan} {et~al.}(2013){Ninan}, {Ojha}, {Bhatt}, {Ghosh}, {Mohan},
  {Mallick}, {Tamura}, \& {Henning}}]{ninan2013}
{Ninan}, J.~P., {Ojha}, D.~K., {Bhatt}, B.~C., {et~al.} 2013, \apj, 778, 116

\bibitem[{{Ninan} {et~al.}(2015){Ninan}, {Ojha}, {Baug}, {Bhatt}, {Mohan},
  {Ghosh}, {Men'shchikov}, {Anupama}, {Tamura}, \& {Henning}}]{ninan2015}
{Ninan}, J.~P., {Ojha}, D.~K., {Baug}, T., {et~al.} 2015, \apj, 815, 4

\bibitem[{{Principe} {et~al.}(2018){Principe}, {Cieza}, {Hales}, {Zurlo},
  {Williams}, {Ru{\'\i}z-Rodr{\'\i}guez}, {Canovas}, {Casassus},
  {Mu{\v{z}}i{\'c}}, {Perez}, {Tobin}, \& {Zhu}}]{principe2018}
{Principe}, D.~A., {Cieza}, L., {Hales}, A., {et~al.} 2018, \mnras, 473, 879

\bibitem[{{Prusti} {et~al.}(1993){Prusti}, {Bontekoe}, {Chiar}, {Kester}, \&
  {Whittet}}]{prusti1993}
{Prusti}, T., {Bontekoe}, T.~R., {Chiar}, J.~E., {Kester}, D.~J.~M., \&
  {Whittet}, D.~C.~B. 1993, \aap, 279, 163

\bibitem[{{Reipurth}(1985)}]{reipurth1985}
{Reipurth}, B. 1985, \aap, 143, 435

\bibitem[{{Reipurth} {et~al.}(1997){Reipurth}, {Olberg}, {Gredel}, \&
  {Booth}}]{reipurth1997}
{Reipurth}, B., {Olberg}, M., {Gredel}, R., \& {Booth}, R.~S. 1997, \aap, 327,
  1164

\bibitem[{{Reipurth} \& {Wamsteker}(1983)}]{reipurth1983b}
{Reipurth}, B., \& {Wamsteker}, W. 1983, \aap, 119, 14

\bibitem[{{Reipurth} {et~al.}(1991){Reipurth}, {Brand}, {Wouterloot}, {Herbig},
  {Pettersson}, {Schwartz}, {Nyman}, {Krautter}, {Graham}, {Wilking}, \&
  {Eiroa}}]{reipurth1991}
{Reipurth}, B., {Brand}, J., {Wouterloot}, J.~G.~A., {et~al.} 1991, European
  Southern Observatory Scientific Report, 11, 1

\bibitem[{{Saito} {et~al.}(2012){Saito}, {Hempel}, {Minniti}, {Lucas},
  {Rejkuba}, {Toledo}, {Gonzalez}, {Alonso-Garc{\'{\i}}a}, {Irwin},
  {Gonzalez-Solares}, {Hodgkin}, {Lewis}, {Cross}, {Ivanov}, {Kerins},
  {Emerson}, {Soto}, {Am{\^o}res}, {Gurovich}, {D{\'e}k{\'a}ny}, {Angeloni},
  {Beamin}, {Catelan}, {Padilla}, {Zoccali}, {Pietrukowicz}, {Moni Bidin},
  {Mauro}, {Geisler}, {Folkes}, {Sale}, {Borissova}, {Kurtev}, {Ahumada},
  {Alonso}, {Adamson}, {Arias}, {Bandyopadhyay}, {Barb{\'a}}, {Barbuy},
  {Baume}, {Bedin}, {Bellini}, {Benjamin}, {Bica}, {Bonatto}, {Bronfman},
  {Carraro}, {Chen{\`e}}, {Clari{\'a}}, {Clarke}, {Contreras}, {Corvill{\'o}n},
  {de Grijs}, {Dias}, {Drew}, {Fari{\~n}a}, {Feinstein},
  {Fern{\'a}ndez-Laj{\'u}s}, {Gamen}, {Gieren}, {Goldman},
  {Gonz{\'a}lez-Fern{\'a}ndez}, {Grand}, {Gunthardt}, {Hambly}, {Hanson},
  {He{\l}miniak}, {Hoare}, {Huckvale}, {Jord{\'a}n}, {Kinemuchi}, {Longmore},
  {L{\'o}pez-Corredoira}, {Maccarone}, {Majaess}, {Mart{\'{\i}}n}, {Masetti},
  {Mennickent}, {Mirabel}, {Monaco}, {Morelli}, {Motta}, {Palma}, {Parisi},
  {Parker}, {Pe{\~n}aloza}, {Pietrzy{\'n}ski}, {Pignata}, {Popescu}, {Read},
  {Rojas}, {Roman-Lopes}, {Ruiz}, {Saviane}, {Schreiber}, {Schr{\"o}der},
  {Sharma}, {Smith}, {Sodr{\'e}}, {Stead}, {Stephens}, {Tamura}, {Tappert},
  {Thompson}, {Valenti}, {Vanzi}, {Walton}, {Weidmann}, \&
  {Zijlstra}}]{saito2012}
{Saito}, R.~K., {Hempel}, M., {Minniti}, D., {et~al.} 2012, \aap, 537, A107

\bibitem[{{Scoville} {et~al.}(1980){Scoville}, {Krotkov}, \&
  {Wang}}]{scoville1980}
{Scoville}, N.~Z., {Krotkov}, R., \& {Wang}, D. 1980, \apj, 240, 929

\bibitem[{{Smette} {et~al.}(2015){Smette}, {Sana}, {Noll}, {Horst}, {Kausch},
  {Kimeswenger}, {Barden}, {Szyszka}, {Jones}, {Gallenne}, {Vinther},
  {Ballester}, \& {Taylor}}]{smette2015}
{Smette}, A., {Sana}, H., {Noll}, S., {et~al.} 2015, \aap, 576, A77

\bibitem[{{Smith}(1995)}]{smith1995}
{Smith}, M.~D. 1995, \aap, 296, 789

\bibitem[{{Takami} {et~al.}(2006){Takami}, {Chrysostomou}, {Ray}, {Davis},
  {Dent}, {Bailey}, {Tamura}, {Terada}, \& {Pyo}}]{takami2006}
{Takami}, M., {Chrysostomou}, A., {Ray}, T.~P., {et~al.} 2006, \apj, 641, 357

\bibitem[{{Turner} {et~al.}(1977){Turner}, {Kirby-Docken}, \&
  {Dalgarno}}]{turner1977}
{Turner}, J., {Kirby-Docken}, K., \& {Dalgarno}, A. 1977, The Astrophysical
  Journal Supplement Series, 35, 281

\bibitem[{{Udalski} {et~al.}(2015){Udalski}, {Szyma{\'n}ski}, \&
  {Szyma{\'n}ski}}]{udalski2015}
{Udalski}, A., {Szyma{\'n}ski}, M.~K., \& {Szyma{\'n}ski}, G. 2015, \actaa, 65,
  1

\bibitem[{{Vernet} {et~al.}(2011){Vernet}, {Dekker}, {D'Odorico}, {Kaper},
  {Kjaergaard}, {Hammer}, {Randich}, {Zerbi}, {Groot}, {Hjorth}, {Guinouard},
  {Navarro}, {Adolfse}, {Albers}, {Amans}, {Andersen}, {Andersen}, {Binetruy},
  {Bristow}, {Castillo}, {Chemla}, {Christensen}, {Conconi}, {Conzelmann},
  {Dam}, {de Caprio}, {de Ugarte Postigo}, {Delabre}, {di Marcantonio},
  {Downing}, {Elswijk}, {Finger}, {Fischer}, {Flores}, {Fran{\c{c}}ois},
  {Goldoni}, {Guglielmi}, {Haigron}, {Hanenburg}, {Hendriks}, {Horrobin},
  {Horville}, {Jessen}, {Kerber}, {Kern}, {Kiekebusch}, {Kleszcz}, {Klougart},
  {Kragt}, {Larsen}, {Lizon}, {Lucuix}, {Mainieri}, {Manuputy}, {Martayan},
  {Mason}, {Mazzoleni}, {Michaelsen}, {Modigliani}, {Moehler}, {M{\o}ller},
  {Norup S{\o}rensen}, {N{\o}rregaard}, {P{\'e}roux}, {Patat}, {Pena}, {Pragt},
  {Reinero}, {Rigal}, {Riva}, {Roelfsema}, {Royer}, {Sacco}, {Santin},
  {Schoenmaker}, {Spano}, {Sweers}, {Ter Horst}, {Tintori}, {Tromp}, {van
  Dael}, {van der Vliet}, {Venema}, {Vidali}, {Vinther}, {Vola}, {Winters},
  {Wistisen}, {Wulterkens}, \& {Zacchei}}]{vernet2011}
{Vernet}, J., {Dekker}, H., {D'Odorico}, S., {et~al.} 2011, \aap, 536, A105

\bibitem[{{Vorobyov} \& {Basu}(2006)}]{vorobyov2006}
{Vorobyov}, E.~I., \& {Basu}, S. 2006, \apj, 650, 956

\bibitem[{{Vorobyov} \& {Basu}(2010)}]{vorobyov2010}
---. 2010, \apj, 719, 1896

\bibitem[{{Vorobyov} {et~al.}(2013){Vorobyov}, {Zakhozhay}, \&
  {Dunham}}]{vorobyov2013}
{Vorobyov}, E.~I., {Zakhozhay}, O.~V., \& {Dunham}, M.~M. 2013, \mnras, 433,
  3256

\bibitem[{{Wils} {et~al.}(2009){Wils}, {Greaves}, {Catelan}, {Djorgovski},
  {Drake}, {Graham}, {Mahabal}, {Williams}, {Beshore}, {Larson}, \&
  {Christensen}}]{wils2009}
{Wils}, P., {Greaves}, J., {Catelan}, M., {et~al.} 2009, The Astronomer's
  Telegram, 2307, 1

\bibitem[{{Wolniewicz} {et~al.}(1998){Wolniewicz}, {Simbotin}, \&
  {Dalgarno}}]{wolniewicz1998}
{Wolniewicz}, L., {Simbotin}, I., \& {Dalgarno}, A. 1998, The Astrophysical
  Journal Supplement Series, 115, 293

\bibitem[{{Wright} {et~al.}(2010){Wright}, {Eisenhardt}, {Mainzer}, {Ressler},
  {Cutri}, {Jarrett}, {Kirkpatrick}, {Padgett}, {McMillan}, {Skrutskie},
  {Stanford}, {Cohen}, {Walker}, {Mather}, {Leisawitz}, {Gautier}, {McLean},
  {Benford}, {Lonsdale}, {Blain}, {Mendez}, {Irace}, {Duval}, {Liu}, {Royer},
  {Heinrichsen}, {Howard}, {Shannon}, {Kendall}, {Walsh}, {Larsen}, {Cardon},
  {Schick}, {Schwalm}, {Abid}, {Fabinsky}, {Naes}, \& {Tsai}}]{wright2010}
{Wright}, E.~L., {Eisenhardt}, P.~R.~M., {Mainzer}, A.~K., {et~al.} 2010, \aj,
  140, 1868

\end{thebibliography}



\end{document}